%% file: Roversi-08-walt-srncompl.tex
\documentclass{article}
\include{environment}

\linenumbers
\begin{document}
\include{front-matter}
\include{introduction}
\include{intuitions}
\include{overview-WALT}
\include{safe-recursion-on-notation}
\include{square-composition}
\include{completeness}
\include{conclusions}
\bibliographystyle{alpha}
\bibliography{Roversi-08-walt-completeness}
\appendix
\input{proofs}
\end{document}

%% file: environment.tex
\usepackage{pstricks}
\usepackage[modulo]{lineno}
\usepackage{amsmath}
\usepackage{stmaryrd}
\usepackage{proof}
\usepackage{epsfig}
\usepackage{subfigure}
\usepackage{pxfonts} 
\usepackage[all]{xy} %

\newcommand{\tc}[2]{\textcolor{#1}{#2}}   
\newcommand{\PT}{$\Lambda$} 
\newcommand{\PTT}{$\Lambda^{\texttt{T}}$} 
\newcommand{\SF}{\textsf{System F}}
\newcommand{\ST}{\textsf{System T}}
\newcommand{\ICC}{\textsf{ICC}}
\newcommand{\FP}{\textsf{FP}}

\newcommand{\HOLRR}{\textsf{HOLRR}}
\newcommand{\BC}{\textsf{BC}}
\newcommand{\BNS}{\textsf{HTRR}}
\newcommand{\ILAL}{\textsf{ILAL}}
\newcommand{\LLL}{\textsf{LLL}}

\newcommand{\WALT}{\textsf{WALT}}
\newcommand{\SRN}{\textsf{SRN}}
\newcommand{\RlSRN}{\textsf{RlSRN}}
\newcommand{\QlSRN}{\textsf{ClSRN}}
\newcommand{\SRNVnames}{\textsf{V}}
\newcommand{\llan}{\langle\!\langle}
\newcommand{\rran}{\rangle\!\rangle}
\newcommand{\dpth}[1]{\operatorname{d}({#1})}  
\newcommand{\nf}[1]{\operatorname{nf}(#1)}
\newcommand{\dom}[1]{\operatorname{Dom}(#1)}
\newcommand{\size}[1]{|#1|}      
\newcommand{\nocc}[2]{\operatorname{no}({#1, #2})}
\newcommand{\PTV}{$\Lambda_{\texttt{V}}$} 
\newcommand{\FV}[1]{\operatorname{FV}(#1)}              
\newcommand{\subs}[2]{\{^{#1}/_{#2}\}}  %
\newcommand{\seq}[3]{\vec{#1}_{[#2;#3]}}
\newcommand{\Nat}{\mathbb{N}}
\newcommand{\bs}{\backslash}
\newcommand{\lan}{\langle}
\newcommand{\ran}{\rangle}
\newcommand{\elan}{\langle\!\lbrace}
\newcommand{\eran}{\rbrace\!\rangle}
\newcommand{\ta}[2]{#1\!:\!#2}
\newcommand{\rew}{\leadsto_{w}}      
\newcommand{\BNum}[1]{\overline{\overline{#1}}}
\newcommand{\BSuccZ}{\mathtt{Ws0}}
\newcommand{\BSuccO}{\mathtt{Ws1}}
\newcommand{\UNum}[1]{\overline{#1}}
\newcommand{\Intg}[1]{\mathsf{#1}}
\newcommand{\BIntT}{\mathbf{W}}
\newcommand{\UIntT}{\mathbf{N}}
\newcommand{\ListT}{\mathbf{L}}
\newcommand{\li}{\multimap}   %
\newcommand{\liv}{\multimapdot}
\newcommand{\Zero}{0}                        
\newcommand{\zero}[2]{\mathtt{z}^{#1;#2}}    
\newcommand{\Suc}[1]{\mathtt{s}_{#1}}        
\newcommand{\Sucz}{\mathtt{s}_0}             
\newcommand{\Suco}{\mathtt{s}_1}             
\newcommand{\Pred}{\mathtt{P}}               
\newcommand{\sucz}{\mathtt{s}^{0;1}_0}       
\newcommand{\suco}{\mathtt{s}^{0;1}_1}       
\newcommand{\preds}{\mathtt{p}}              
\newcommand{\pred}{\mathtt{p}^{0;1}}         
\newcommand{\proj}[3]{\pi^{#1;#2}_{#3}}      
\newcommand{\Branch}{\mathtt{B}}
\newcommand{\bran}{\mathtt{c}^{0,3}}         
\newcommand{\rec}[3]{\mathtt{r}^{#1;#2}[#3]} 
\newcommand{\comp}[5]{\circ^{#1;#2}_{#3;#4}[#5]} 
\newcommand{\lcomp}[5]{\diamond^{#1;#2}_{#3;#4}[#5]} 
\newcommand{\share}[3]{\mathtt{Y}^{#1;#2}_{#3}} 
\newcommand{\mshare}[4]{\mathtt{mY}^{#1;(#2,#3)}_{#4}} 
\newcommand{\rmshare}[4]{\circlearrowleft\!\!\mathtt{mY}^{#1;(#2,#3)}_{#4}} 
\newcommand{\rotate}[3]{\circlearrowleft^{#1;#2}_{#3}}  
\newcommand{\Iter}[5]{\mathtt{It}_{#1; #2}[#3,#4,#5]} 
\newcommand{\sqcomp}[3]{\boxdot^{#1;#2}_{#3}} 
\newcommand{\mshsqcomp}[4]{\curlyvee^{#1;#2\bs#3}_{#4}} 
\newcommand{\BEmbed}[2]{\mathtt{Eb}^{#1}[#2]}
\newcommand{\LEmbed}[3]{\mathtt{El}^{#1}_{#2}[#3]}
\newcommand{\EEmbed}[4]{\mathtt{Ee}^{#1}_{#2;#3}[#4]}
\newcommand{\Coerc}{\mathtt{Coerce}}
\renewcommand{\S}{\$}
\newcommand{\srtw}[2]{\llbracket#1\rrbracket_{#2}} 
\newcommand{\etp}[1]{[#1]^{\circ}} 
\newcommand{\wght}[2]{\operatorname{wg}_{#2}(#1)}   

\newtheorem{theorem}{Theorem}[section]

\newtheorem{corollary}[theorem]{Corollary}

%% file: front-matter.tex
\pagestyle{headings}
\title{Weak Affine Light Typing is complete\\ with respect to Safe Recursion on Notation}
\author{Luca Roversi
\footnote{Dipartimento di Informatica,
          Universit\`a di Torino,
          C.so Svizzera n.ro 185 --- 10149, TORINO --- ITALY.
}
\footnote{
\textit{e-mail}:\texttt{roversi@di.unito.it}.
\textit{home page}:\texttt{http://www.di.unito.it/\~{}rover}.
}
}
\date{}
\maketitle
\input{abstract}
\tableofcontents

%% file: abstract.tex
\begin{abstract}
Weak affine light typing (\WALT) assigns light affine linear formulae as types to a subset of $\lambda$-terms of \SF. \WALT\ is poly-time sound: if a $\lambda$-term $M$ has type in \WALT, $M$ can be evaluated with a polynomial cost in the dimension of the derivation that gives it a type. The evaluation proceeds under any strategy of a rewriting relation which is a mix of both call-by-name and call-by-value $\beta$-reductions.
\WALT\ \textit{weakens}, namely \textit{generalizes}, the notion of ``\textit{stratification of deductions}'', common to some \textsf{Light Systems} --- those logical systems, derived from Linear logic, to characterize the set of Polynomial functions --- .
A weaker stratification allows to define a compositional \textit{embedding} of Safe recursion on notation (\SRN) into \WALT. It turns out that the expressivity of \WALT\ is strictly stronger than the one of the known \textsf{Light Systems}.
The embedding passes through the representation of a subsystem of \SRN. It is obtained by restricting the composition scheme of \SRN\ to one that can only use its safe variables \textit{linearly}. On one side, this suggests that \SRN, in fact, can be redefined in terms of more primitive constructs. On the other,
the \textit{embedding} of \SRN\ into \WALT\ enjoys the two following remarkable aspects. Every datatype, required by the embedding, is represented from scratch, showing the strong structural proof-theoretical roots of \WALT.
Moreover, the embedding highlights a stratification structure of the normal and safe arguments, normally hidden inside the world of \SRN-normal/safe variables: the less an argument is ``polyomially impredicative'', the deeper, in a formal, proof-theoretical sense, it is represented inside \WALT. Finally, since \WALT\ is \SRN-complete it is also polynomial-time complete since \SRN\ is.
\end{abstract}
\newpage

%% file: introduction.tex
\section{Introduction}
\label{section:Introduction}
Implicit computational complexity (\ICC) explores machine-independent characterizations of complexity classes without any explicit reference to resource usage bounds, which, instead, result from restricting suitable computational structures. Contributions to \ICC\ can have their roots in the recursion theory \cite{Cobham65,Bellantoni92CC,Leivant94RRII,Leivant95RRI,Leivant99RRIII,LeivantMarion00RRIV}, in the structural proof-theory and linear logic \cite{Girard:1998-IC,Lafont02SLL}, in the rewriting systems or functional programming \cite{Huet:JACM-80,Dershowitz:TCS-92,Jones:TCS-99,Leivant93RR,Leivant94ic}, in the type systems \cite{hofmann97csl,hofmann99linear,hofmann99thesis,hofmann00safe,Bellantoni00APAL,Bellantoni01MSS}\ldots.
\par
One specific goal of \ICC\ is to make evident that the known complexity classes are concepts with an intrinsic mathematical nature. A way of achieving the goal is to formally relate the known \ICC\ characterizations.
\par
Here, we accomplish the goal relatively to two \ICC\ systems that characterize the class \FP\ of \textit{Polynomial functions}.
Specifically, we formally relate the basic concept of ``predicative recursion'' of Safe recursion on notation (\SRN) \cite{Bellantoni92CC}, and the notion of ``stratification of the derivations'', basic for Light linear logic (\LLL) \cite{Girard:1998-IC} and for various proof/type theoretical systems, derived from it.
The strategy to formalize the relation is to embed, inductively, \SRN\ into Weak affine light typing (\WALT) \cite{Roversi:2007-WALT-FULL}, a type system for pure $\lambda$-terms that strictly generalizes Intuitionistic light affine logic (\ILAL) \cite{Asperti:1998-LICS,Asperti02TOCL,Terui:2001-LICS,Terui:2002-AML}.
\par
Recall that \SRN\ is a recursion theoretical system.
It is generated from a set of basic functions, closed under the \textit{safe composition} and \textit{safe recursion} schemes. \SRN\ captures \FP\ by partitioning the set of arguments of each function $g(\vec{n},\vec{s})$ into those that are \textit{normal}, namely $\vec{n}$, and those that are \textit{safe}, \textit{i.e.} $\vec{s}$. The basic functions can only have safe arguments. The crucial features of a function $f$, defined by an instance of the safe recursive scheme, are: (i) the unfolding of $f$ is driven by a normal argument, and (ii) the recursive call of $f$ may only appear in a safe argument position, as far as the unfolding proceeds, ensuring that recursion over the result of a function defined by recursion is not possible. Section~\ref{section:Safe Recursion on Notation} formally recalls \SRN. In fact, it also recalls \textit{Composition-linear safe recursion on notation} (\QlSRN), already introduced in \cite{Roversi:2007-WALT-FULL} where, however, it was called as \textit{Quasi-linear safe recursion on notation}. \QlSRN\ restricts \SRN; It is defined on a set of basic functions, closed under the \textit{full} safe recursion scheme of \SRN\ and a \textit{linear safe composition scheme} that uses linearly its safe variables. Namely, \QlSRN\ strictly generalizes $\BC^-$ \cite{MurawskiOng00}. Recall that $\BC^-$ is \SRN\ where \textit{both} the safe composition and recursion schemes can exclusively use their safe variables \textit{linearly}.
\par
The reason to recall \QlSRN\ here is the following one. We already know that \QlSRN\ can be embedded into \WALT\ \cite{Roversi:2007-WALT-FULL}. By using that result, we show that, in fact, \textit{full} \SRN\ can be embedded into \WALT, so relating ``predicative recursion'' to ``stratification'' without any restriction.
\par
Formally, the relation reads as follows. There exists an interpretation map $\srtw{\,}{}$ from \SRN\ to \WALT, such that, for every $f(n_1,\ldots,n_k,s_1,\ldots,s_l)\in\SRN$, with $k$ \textit{normal} and $l$ \textit{safe} arguments, we can prove that:
(i) if $f(n_1,\ldots,n_k,s_1,\ldots,s_l)=n$, then
$\srtw{f(n_1,\ldots,n_k,s_1,\ldots,s_l)}{}$ reduces to $\srtw{n}{}$, using $\rew$, a rewriting relation which is a mix of both call-by-name and call-by-value $\beta$-reductions, and
(ii) $\srtw{f(n_1,\ldots,n_k,s_1,\ldots,s_l)}{}$ has type $\$^{m}\BIntT$, since
$\srtw{f}{}$ has type $\overbrace{\$\BIntT\liv\ldots\liv\$\BIntT}^{k}
 \liv
 \overbrace{\$^{m}\BIntT\liv\ldots\liv\$^{m}\BIntT}^{l}\liv\$^{m}\BIntT$,
for some $m\geq 1$, the type $\BIntT$ being the one for binary words in \WALT.
\par
Point (i) shows that \WALT\ is \SRN\textit{-complete}. Namely, \WALT\ is the first system, derived as a restriction of Linear logic to characterize \FP, where \textit{full} \SRN\ can be embedded.
\par
Point (ii) links $m$ to the complexity of the definition of $f$: $m$ depends on the number of nested safe compositions and of safe recursive schemes that define $f$. The types explicitly show a layered structure inside the normal and safe arguments of \SRN: the type of a safe argument is $m\geq 1$ $\$$-modality occurrences deep because a safe argument can be used in the course of a recursive unfolding to produce a result. Orthogonally, the depth of the type of every normal argument is limited to $1$, so giving to every normal argument the necessary ``replication power'', required to duplicate syntactic structure in the course of an unfolding.
On one side, this means that the Light linear logic-like systems say that the weaker is the possibility of a word to replicate structure, behaving it as an iterator, the deeper is its type. On the other, the recursive systems like \SRN\ are based exactly on the reversed idea, though this cannot be formally stated in terms of any typing information inside \SRN.
\par
A further consequence of embedding \SRN\ into \WALT\ is that we obtain a second version of such a proof, the first being in \cite{Roversi:2007-WALT-FULL}. Recall that ``polytime completeness'' means that every polynomial Turing machine can be represented as a term of \WALT.
\par
However, a more relevant consequence of looking for the formal relations between two systems like \SRN\ and \WALT, as we have just done, is the way we prove the \SRN-completeness of \WALT. We shall see that it is obtained by \textit{simulating the full composition scheme} of \SRN\ through the \textit{linear} safe composition scheme of \QlSRN\ and its \textit{full} recursive scheme, which coincides to the one of \SRN. This candidates \QlSRN\ to be \textit{a linear kernel of} \SRN, so pointing to the existence of a reformulation of \SRN\ itself in terms of more primitive and linear constructs, as we shall discuss in the conclusions.
\par
Finally, we observe that \WALT\ yields a higher-order characterization of \FP\ in the lines of Higher type ramified recursion (\BNS) \cite{Bellantoni00APAL}  and Higher order linear ramified recursion (\HOLRR) \cite{DalLago+Martini+Roversi:2004-TYPES}, but with a relevant difference.
Both \BNS\ and \HOLRR\ build their terms by assuming the existence of constant symbols, like words, successors, \textit{etc.}. On the contrary, no constant symbol is used inside \WALT\ where everything is defined from scratch, exploiting its II-order structural proof-theoretic roots.
\paragraph{Outline.}
Section~\ref{section:Overviewing WALT intuitively} intuitively recalls the main intuitions about \WALT, by pointing out how it weakens the design principles of \ILAL.
Section~\ref{section:Overviewing WALT technically} recalls the technical parts of \WALT, required to program the combinators that allow to represent the full composition scheme of \SRN\ in \WALT, as shown in Section~\ref{section:The safe composition in WALT}, so yielding the \SRN-completeness.
Section~\ref{section:Safe Recursion on Notation} formally recalls \SRN\ in the style of \cite{Beckmann:96-AML}.
Section~\ref{section:Conclusions and future work} delineates some possible research directions.

%% file: intuitions.tex
\section{Overviewing \WALT\ intuitively}
\label{section:Overviewing WALT intuitively}
The full technical introduction of \WALT\ is \cite{Roversi:2007-WALT-FULL}.
\par
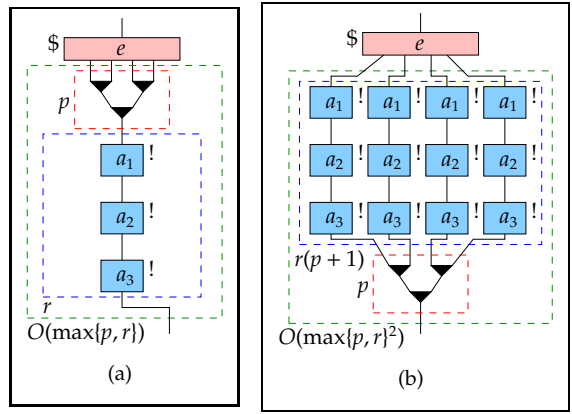
\begin{figure}[ht]
\begin{center}
\begin{tabular}{ccc}
\begin{minipage}[c]{2.5cm}
\fbox{
\subfigure[]{\scalebox{.7}{\input{squaring-chain.pstex_t}}}
}
\end{minipage}
&
&
\begin{minipage}[c]{4cm}
\fbox{
\subfigure[]{\scalebox{.7}{\input{squaring-chain-reduced.pstex_t}}}
}
\end{minipage}
\end{tabular}
\end{center}
\caption{A squaring chain and its reduct}
\label{caption:squaring chain}
\end{figure}
\begin{figure}[ht]
\begin{center}
\begin{tabular}{cp{.5cm}c}
\begin{minipage}[c]{5.5cm}
\fbox{
\subfigure[]{\scalebox{.7}{\input{weak-squaring-chain.pstex_t}}
}
}
\end{minipage}
&&
\begin{minipage}[c]{4cm}
\fbox{
\subfigure[]{\scalebox{.7}{\input{weak-squaring-chain-to-squaring.pstex_t}}}
}
\end{minipage}
\end{tabular}
\end{center}
\caption{A weak squaring and a stuck chain}
\label{caption:Weak squaring chain}
\end{figure}
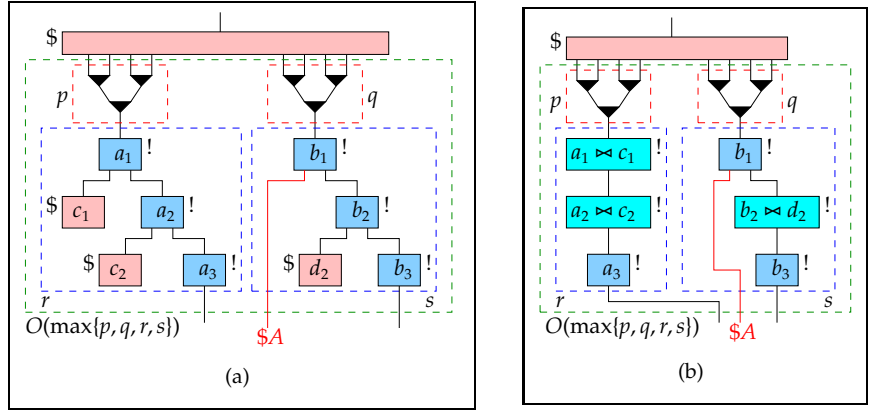

Here we want to recall the key ideas about \WALT\ at an intuitive level. The main goal is to illustrate the main reasons why \WALT\ is more expressive than other deductive systems, derived from Linear logic, to characterize the class of polynomial functions (\FP).
\paragraph{Squaring chains.}
\WALT\ contains the main ``complex'' structure of Intuitionistic light affine logic (\ILAL) which we call \textit{squaring chain}. A graph representation of an instance of squaring chain is in Figure~\ref{caption:squaring chain}.(a).
On top of it there is a tree of nodes, all the black triangles, that contract a set of assumptions of the given topmost $\$$-box $e$. Below the tree of contraction nodes there is a list of $!$-boxes $a_1, a_2, a_3$. Every of them depends on at most a single assumption, which is the basic constraint of $!$-boxes of \ILAL. Of course, generally, the number of contractions nodes and of $!$-boxes in a squaring chain is arbitrary and the tree they form need not to be perfectly balanced.
The chain is dubbed as ``squaring'' because its normalization leads to the configuration in Figure~\ref{caption:squaring chain}.(b), where the size $O(\max\{p,r\})$, essentially, ``squares'' to a value which is $O(\max\{p,r\}^2)$.
\paragraph{Weak squaring and stuck chains.}
However, besides the squaring chains, \WALT\ contains both \textit{weak squaring chains} and \textit{stuck chains}, and its expressive power relies on their existence. A graph representation of both types of chains is in
Figure~\ref{caption:Weak squaring chain}.(a).
The one to the left is a weak squaring chain. The other, to the right, is a stuck chain.
\par
\begin{figure}[ht]
\begin{center}
\begin{tabular}{cc}
\begin{minipage}[c]{7cm}
\fbox{
\scalebox{.7}{\input{weak-squaring-chain-reduced.pstex_t}
}
}
\end{minipage}
\end{tabular}
\end{center}
\caption{The reduct of a weak squaring chain}
\label{caption:Weak squaring chain reduced}
\end{figure}
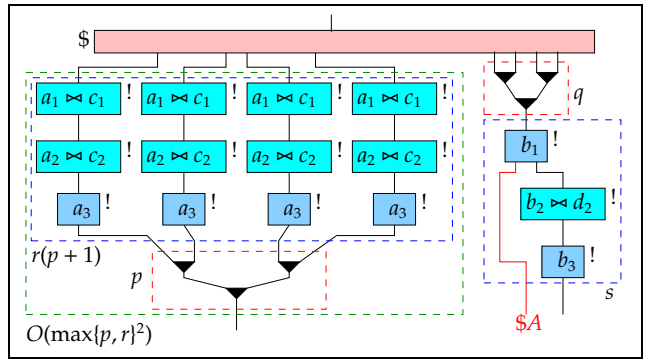

The weak squaring chain contains a tree of contraction nodes. However, it is based on a more liberalized form of $!$-boxes. They may depend on more than one assumption: one of them must have a $!$-modal type, the others a $\$$-modal type. The chain under description is ``weak squaring'' because \textit{only after} we merge the $\$$-boxes $c_1, c_2$ into the $!$-boxes $a_1, a_2$, respectively, it transforms to a squaring chain, with $!$-boxes $a_1\bowtie c_1, a_2\bowtie c_2, a_3$, that can be squared to the configuration to the left in Figure~\ref{caption:Weak squaring chain reduced}. We insist on observing that, before the merging of boxes, no squaring can occur.
\par
This is why the configuration to the right in Figure~\ref{caption:Weak squaring chain}.(b) is a stuck chain, and not a weak squaring one.
Its ``squaring through normalization'' cannot start, even if we merge $b_2$ and $d_2$, because there is no $\$$-box plugged into the assumption of type $\$A$ of the $!$-box $b_1$.
So, the chain to the right in Figure~\ref{caption:Weak squaring chain}.(b) is stuck until the context, eventually, supplies a closed $\$$-box with conclusion $\$A$ that can be merged into $b_1$, so yielding a squaring chain that we can ``square''.
\par
We conclude by remarking that the form of weak squaring and stuck chains is more general than the one in the given example. Indeed, not only a single closed $\$$-box can be dangling down the $\$$-modal assumptions of a $!$-box, but there can be a whole tree whose nodes can only be $\$$-boxes, which must be closed whenever they constitute the leaves of the tree itself.
\paragraph{The lazy nature of \WALT.}
\WALT\ induces a call-by-value dynamics on the $\lambda$-terms it gives types to as consequence of the more general form of its $!$-boxes, as compared to \ILAL. As we have seen, a chain is stuck until the context supplies the $\$$-boxes that close all the assumptions, with $\$$-modal type, of those $!$-boxes that must be duplicated. All such assumptions are used to represent the parameters in the simulation of the full recursive scheme of \SRN\ inside \WALT, with the right type. We shall recall the main idea under some simplifying assumptions, to keep things more readable.
\par
Let us assume to have a function $f$, recursively defined as $f(0,a) = g(a)$, and $f(n,a) = h(n-1,a,f(n-1,a))$, with $n\geq1$. We want to show how simulating its top-down recursive unfolding:
\small
\begin{align*}
f(n,a) =  h(n-1,a,f(n-1,a))
= \ldots =  h(n-1,a,h(n-2,a,\ldots h(0,a,g(a))\ldots))
\end{align*}
\normalsize
by a bottom-up reconstruction that iterates some transition functions on suitable configurations and pre-configurations.
The reconstruction requires to assume $H, G$ be the interpretations of $h, g$, respectively, in \WALT. Moreover, we assume
the unary strings $\UNum{n}, \UNum{a}$, not words, represent $n, a$ in \WALT.
What we are going to say, though, keeps holding with $f$ of arbitrary arity and with words as its arguments, instead of strings.
In \WALT\ we can develop sequences of computations like the following one, where all the terms can be correctly typed:
\small
\begin{align}
\label{align:intro-step1}
&
\llan  G\UNum{a}
        , [\underbrace{\UNum{0},\ldots,\UNum{0}}_{n+1}]
        , [\underbrace{\UNum{a},\ldots,\UNum{a}}_{n+1}] \rran
\rew^{*}
\\
\label{align:intro-step2}
\llan G\UNum{a}
     ,\lan \UNum{0},\underbrace{[\UNum{1},\ldots,\UNum{1}]}_{n}\ran
     ,\lan \UNum{a},\underbrace{[\UNum{a},\ldots,\UNum{a}]}_{n}    \ran     \rran
\rew^{*}
&
\llan H\,\UNum{0}\,\UNum{a}\,(G\UNum{a})
    ,\underbrace{[\UNum{1},\ldots,\UNum{1}]}_{n}
    ,\underbrace{[\UNum{a},\ldots,\UNum{a}]}_{n} \rran
\rew^{*}
\\
\label{align:intro-step3}
\llan H\,\UNum{0}\,\UNum{a}\,(G\UNum{a})
     ,\lan \UNum{1},\underbrace{[\UNum{2},\ldots,\UNum{2}]}_{n-1}\ran
     ,\lan \UNum{a},\underbrace{[\UNum{a},\ldots,\UNum{a}]}_{n-1}    \ran     \rran
\rew^{*}
&
\llan H\,\UNum{1}\,\UNum{a}\,(H\,\UNum{0}\,\UNum{a}\,(G\UNum{a}))
    ,\underbrace{[\UNum{1},\ldots,\UNum{1}]}_{n-1}
    ,\underbrace{[\UNum{a},\ldots,\UNum{a}]}_{n-1} \rran
\rew^{*}\ldots
\end{align}
\normalsize
The ideal column to the right of $\rew^{*}$ contains \textit{configurations}, the topmost being the \textit{initial} one.
The column to the left of $\rew^{*}$ contains \textit{pre-configurations}.
Every pre-configuration comes from its preceding configuration by
(i) separating head and tail of every list, and storing them as the two components of a same pair,
(ii) only on the leftmost list, simultaneously to the separation, the successor is mapped on the tail.
\par
Every configuration, other than the initial one, is obtained from its preceding pre-configuration by the application of an instance of $H$ to the first element of every pair, and to the first element of the whole pre-configuration, which accumulates the partial result of the bottom-up reconstruction.
\par
\begin{figure}[ht]
\begin{center}
\begin{tabular}{cc}
\fbox{
\begin{minipage}[c]{11.5cm}
\scalebox{.7}{\input{list-of-constants-generation.pstex_t}
}
\end{minipage}
}
\end{tabular}
\end{center}
\caption{The term that generates the list of constants in a recursive scheme}
\label{caption:list of constants in a recursive scheme}
\end{figure}
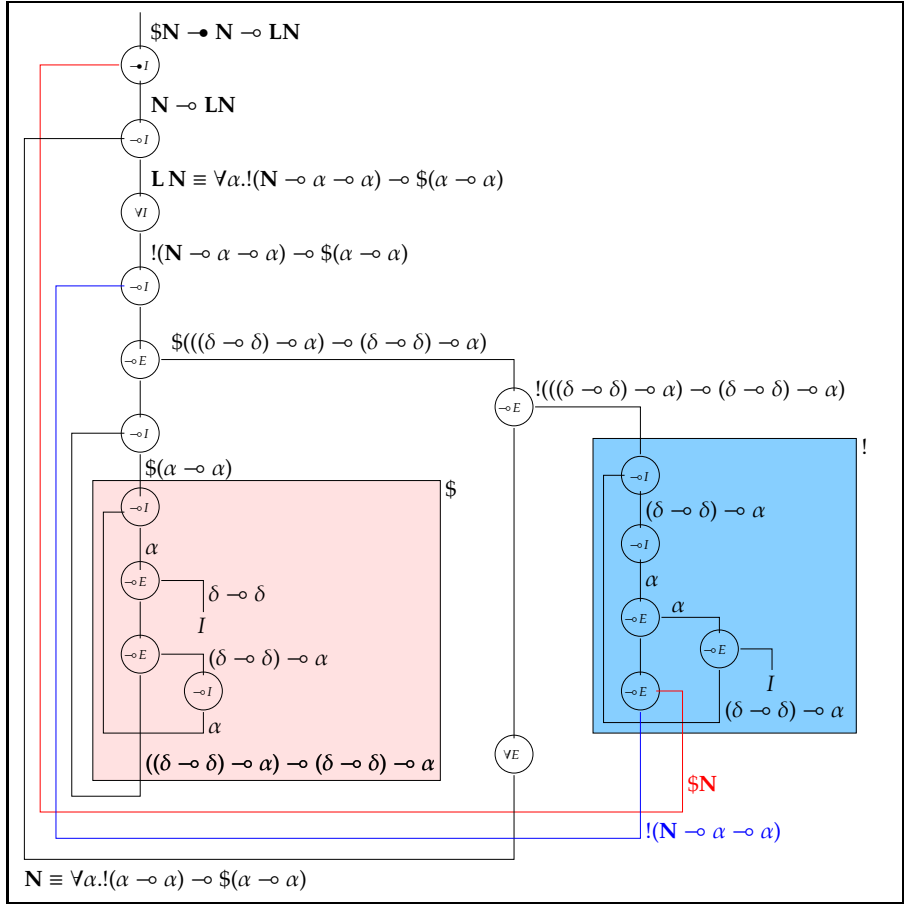
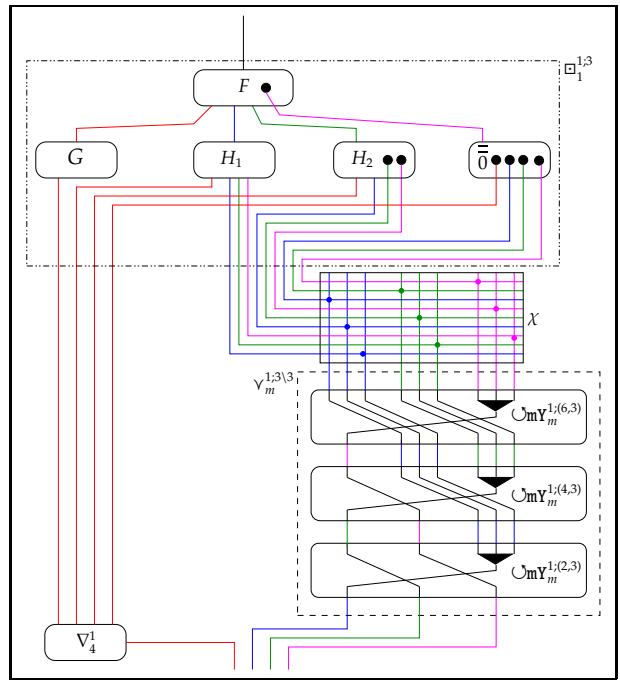
\begin{figure}[ht]
\begin{center}
\begin{tabular}{ccc}
\begin{minipage}[c]{.5\textwidth}
\fbox{
\scalebox{.6}{\input{square-composition-block-diagram.pstex_t}}
}
\end{minipage}
\end{tabular}
\end{center}
\caption{Functional block scheme of \SRN-composition in \WALT}
\label{figure:square-composition-block-diagram}
\end{figure}

The \textit{main point} for everything to work correctly in the above simulation is to produce $[\underbrace{\UNum{a},\ldots,\UNum{a}}_{n+1}]$ with the right type.
This is obtained by using the term in Figure~\ref{caption:list of constants in a recursive scheme} whose definition is substantially based on an instance of the more general $!$-box existing in \WALT, but not in \ILAL.
The assumption of type $\$\UIntT$ in the $!$-box waits for $\overline{a}$, one $\$$-box deep. As soon as this value is supplied, the $!$-box is ready for the duplication by means of the contraction nodes that may be contained in the Church numeral $\overline{n}$ of type
$\forall \alpha.!(\alpha\li\alpha)\li\$(\alpha\li\alpha)$, which is the second argument of the whole term. Once both $\overline{a}$ and $\overline{n}$ have been given, the result of the whole term is a list of copies of $\overline{a}$, whose type is the one we can expect:
$\forall \alpha.!(\$\UIntT\li\alpha\li\alpha)\li\$(\alpha\li\alpha)$, that we shorten as $\ListT\UIntT$.
The use of the assumption with type $\$\UIntT$ in the $!$-box is the key step to obtain a result of type $\ListT\UIntT$ which, somewhat, absorbs the $\$$-box initially around the parameter of $\overline{a}$. Without this merging we could not obtain a representation of the iterator whose safe arguments are at the same depth as the result, as required to represent the full recursion scheme of \SRN\ in \WALT.
\paragraph{The full \SRN-composition scheme in \WALT.}
Once the full recursion scheme of \SRN\ is at hand in \WALT, we can use it to encode also the full composition scheme of \SRN.
Figure~\ref{figure:square-composition-block-diagram} shows an example of functional block diagram that summarizes how the full composition scheme of \SRN\ becomes a term of \WALT.
Let us assume $F, G, H_1, H_2$ be terms of \WALT\ that represent the \SRN\ functions $f, g, h_1, h_2$, respectively, and that we need to compose as follows. $f$ takes one normal and two safe arguments which are supplied by $g$, that depends on a single safe argument, and by $h_1, h_2$. Also, we assume that both $h_1, h_2$ require a single normal argument, but $h_1$ needs three safe arguments, while $h_2$ only one.
The first operation to represent the full composition scheme of \SRN\ in \WALT\ is to generate the terms $F, G, H_1, H_2$.
$H_1$ will be obtained from $h_1$ as the result of an inductive translation, as we might expect.
$H_2$ will be defined from $h_2$ with the same inductive process. However, this would lead to a  term with safe arity $1$. To obtain a term $H_2$ with safe arity $3$, we extend the resulting term to erase two of its three safe arguments: $H_2$ and the two bullets close to it in Figure~\ref{figure:square-composition-block-diagram} represent such a final term.
The same holds for $F$ which must erase its third safe argument. Notice that the safe value it erases is supplied by the dummy function $\overline{0}\!\bullet\!\bullet\!\bullet\!\bullet$, which is constantly equal to $\overline{0}$, after the erasure of all its arguments: one normal, the others safe. $G$, supplying it the normal value to $F$, does not present any problem.
The translation process of the functions being composed, all with the same safe arity, occurs inside the square composition $\sqcomp{1}{3}{1}$: the topmost parameter $1$ is the normal arity of every of the terms being composed, $3$ is the maximal safe arity, namely, the value with respect to which we normalize the terms we generate, and the lowermost $1$ is the normal arity of $F$. So, $\sqcomp{1}{3}{1}[F,G,H_1,H_2]$ has normal arity $1$ and safe arity $9$, since every of the three composed terms will have safe arity $3$. After the normalized composition, the ideal functional block $\chi$ rearranges the safe arguments: the first safe arguments of $H_1, H_2\!\bullet\!\bullet$, and $\overline{\overline{0}}\!\bullet\!\bullet\!\bullet\!\bullet$ are put one closed to the others, and the same is done for the second and third ones. The goal is to share each of the group into a single safe argument. This happens inside the functional block $\mshsqcomp{1}{3}{3}{m}$ which applies, one after the other, three further blocks $\rmshare{1}{6}{3}{m},\rmshare{1}{4}{3}{m},\rmshare{1}{2}{3}{m}$.
The behavior of every of these blocks is to share three safe arguments in input into a single safe argument, and to rotate them so that a new group of safe arguments gets ready for the sharing of its components by means of the subsequent block.
The sharing of the safe arguments is hidden inside the black triangles. Every of them contains two one-step long iterations that share the same safe value in the last two positions of a given term $M$. The following unfolding illustrates the idea about the behavior of one of such one-step long iterations:
\begin{align}
\label{align:unfolding-rmshare}
\share{1}{3}{m}[M]\,n_1\,s_1\,s_2\,s_3
&=M\,n_1\,s_1\,s_2\,s_3\,(\share{1}{3}{m}[M]\,n_1\,s_1\,s_2\,s_3 )\\
\nonumber
&=M\,n_1\,s_1\,s_2\,s_3\,
  ((\bs x_1\,y_1\,y_2\,y_3. y_3)\,n_1\,s_1\,s_2\,s_3)\\
\nonumber
&=M\,n_1\,s_1\,s_2\,s_3\,s_3
\end{align}
We insist remarking that \eqref{align:unfolding-rmshare} gives only the idea of what happens. The definition of $\share{1}{3}{m}$ is based on the iteration term, typeable in \WALT, and will be formally given in Section~\ref{section:The safe composition in WALT}, together with its dynamics. Once the safe arguments of $\sqcomp{1}{3}{1}[F,G,H_1,H_2]$ have been shared we are left with a term waiting for three safe and one normal arguments. The latter is replicated four times by $\nabla^1_4$ that, standardly, iterates a tuple of successors, starting from a tuple of four instances of $\overline{\overline{0}}$. Just to remark it again, the above translation mechanism, can be set for any normal and safe arities of any number of functions in \SRN.

%% file: squaring-chain.pstex_t
\begin{picture}(0,0)%
\includegraphics{squaring-chain.pstex}%
\end{picture}%
\setlength{\unitlength}{4144sp}%
\begingroup\makeatletter\ifx\SetFigFontNFSS\undefined%
\gdef\SetFigFontNFSS#1#2#3#4#5{%
  \reset@font\fontsize{#1}{#2pt}%
  \fontfamily{#3}\fontseries{#4}\fontshape{#5}%
  \selectfont}%
\fi\endgroup%
\begin{picture}(1692,2830)(166,-2069)
\put(946,-1006){\makebox(0,0)[lb]{\smash{{\SetFigFontNFSS{12}{14.4}{\familydefault}{\mddefault}{\updefault}{\color[rgb]{0,0,0}$a_2$}%
}}}}
\put(1216,-961){\makebox(0,0)[lb]{\smash{{\SetFigFontNFSS{12}{14.4}{\familydefault}{\mddefault}{\updefault}{\color[rgb]{0,0,0}$!$}%
}}}}
\put(946,-1501){\makebox(0,0)[lb]{\smash{{\SetFigFontNFSS{12}{14.4}{\familydefault}{\mddefault}{\updefault}{\color[rgb]{0,0,0}$a_3$}%
}}}}
\put(1216,-1456){\makebox(0,0)[lb]{\smash{{\SetFigFontNFSS{12}{14.4}{\familydefault}{\mddefault}{\updefault}{\color[rgb]{0,0,0}$!$}%
}}}}
\put(361,479){\makebox(0,0)[lb]{\smash{{\SetFigFontNFSS{12}{14.4}{\familydefault}{\mddefault}{\updefault}{\color[rgb]{0,0,0}$\$$}%
}}}}
\put(451,-16){\makebox(0,0)[lb]{\smash{{\SetFigFontNFSS{12}{14.4}{\familydefault}{\mddefault}{\updefault}{\color[rgb]{0,0,0}$p$}%
}}}}
\put(316,-1771){\makebox(0,0)[lb]{\smash{{\SetFigFontNFSS{12}{14.4}{\familydefault}{\mddefault}{\updefault}{\color[rgb]{0,0,0}$r$}%
}}}}
\put(946,-511){\makebox(0,0)[lb]{\smash{{\SetFigFontNFSS{12}{14.4}{\familydefault}{\mddefault}{\updefault}{\color[rgb]{0,0,0}$a_1$}%
}}}}
\put(1216,-466){\makebox(0,0)[lb]{\smash{{\SetFigFontNFSS{12}{14.4}{\familydefault}{\mddefault}{\updefault}{\color[rgb]{0,0,0}$!$}%
}}}}
\put(181,-1996){\makebox(0,0)[lb]{\smash{{\SetFigFontNFSS{12}{14.4}{\familydefault}{\mddefault}{\updefault}{\color[rgb]{0,0,0}$O(\max\{p,r\})$}%
}}}}
\put(946,434){\makebox(0,0)[lb]{\smash{{\SetFigFontNFSS{12}{14.4}{\familydefault}{\mddefault}{\updefault}{\color[rgb]{0,0,0}$e$}%
}}}}
\end{picture}%

%% file: squaring-chain-reduced.pstex_t
\begin{picture}(0,0)%
\includegraphics{squaring-chain-reduced.pstex}%
\end{picture}%
\setlength{\unitlength}{4144sp}%
\begingroup\makeatletter\ifx\SetFigFontNFSS\undefined%
\gdef\SetFigFontNFSS#1#2#3#4#5{%
  \reset@font\fontsize{#1}{#2pt}%
  \fontfamily{#3}\fontseries{#4}\fontshape{#5}%
  \selectfont}%
\fi\endgroup%
\begin{picture}(2367,2920)(121,-2294)
\put(721,344){\makebox(0,0)[lb]{\smash{{\SetFigFontNFSS{12}{14.4}{\familydefault}{\mddefault}{\updefault}{\color[rgb]{0,0,0}$\$$}%
}}}}
\put(1036,-691){\makebox(0,0)[lb]{\smash{{\SetFigFontNFSS{12}{14.4}{\familydefault}{\mddefault}{\updefault}{\color[rgb]{0,0,0}$a_2$}%
}}}}
\put(1306,-646){\makebox(0,0)[lb]{\smash{{\SetFigFontNFSS{12}{14.4}{\familydefault}{\mddefault}{\updefault}{\color[rgb]{0,0,0}$!$}%
}}}}
\put(1036,-1186){\makebox(0,0)[lb]{\smash{{\SetFigFontNFSS{12}{14.4}{\familydefault}{\mddefault}{\updefault}{\color[rgb]{0,0,0}$a_3$}%
}}}}
\put(1306,-1141){\makebox(0,0)[lb]{\smash{{\SetFigFontNFSS{12}{14.4}{\familydefault}{\mddefault}{\updefault}{\color[rgb]{0,0,0}$!$}%
}}}}
\put(1036,-196){\makebox(0,0)[lb]{\smash{{\SetFigFontNFSS{12}{14.4}{\familydefault}{\mddefault}{\updefault}{\color[rgb]{0,0,0}$a_1$}%
}}}}
\put(1306,-151){\makebox(0,0)[lb]{\smash{{\SetFigFontNFSS{12}{14.4}{\familydefault}{\mddefault}{\updefault}{\color[rgb]{0,0,0}$!$}%
}}}}
\put(1531,-691){\makebox(0,0)[lb]{\smash{{\SetFigFontNFSS{12}{14.4}{\familydefault}{\mddefault}{\updefault}{\color[rgb]{0,0,0}$a_2$}%
}}}}
\put(1801,-646){\makebox(0,0)[lb]{\smash{{\SetFigFontNFSS{12}{14.4}{\familydefault}{\mddefault}{\updefault}{\color[rgb]{0,0,0}$!$}%
}}}}
\put(1531,-1186){\makebox(0,0)[lb]{\smash{{\SetFigFontNFSS{12}{14.4}{\familydefault}{\mddefault}{\updefault}{\color[rgb]{0,0,0}$a_3$}%
}}}}
\put(1801,-1141){\makebox(0,0)[lb]{\smash{{\SetFigFontNFSS{12}{14.4}{\familydefault}{\mddefault}{\updefault}{\color[rgb]{0,0,0}$!$}%
}}}}
\put(1531,-196){\makebox(0,0)[lb]{\smash{{\SetFigFontNFSS{12}{14.4}{\familydefault}{\mddefault}{\updefault}{\color[rgb]{0,0,0}$a_1$}%
}}}}
\put(1801,-151){\makebox(0,0)[lb]{\smash{{\SetFigFontNFSS{12}{14.4}{\familydefault}{\mddefault}{\updefault}{\color[rgb]{0,0,0}$!$}%
}}}}
\put(2026,-691){\makebox(0,0)[lb]{\smash{{\SetFigFontNFSS{12}{14.4}{\familydefault}{\mddefault}{\updefault}{\color[rgb]{0,0,0}$a_2$}%
}}}}
\put(2296,-646){\makebox(0,0)[lb]{\smash{{\SetFigFontNFSS{12}{14.4}{\familydefault}{\mddefault}{\updefault}{\color[rgb]{0,0,0}$!$}%
}}}}
\put(2026,-1186){\makebox(0,0)[lb]{\smash{{\SetFigFontNFSS{12}{14.4}{\familydefault}{\mddefault}{\updefault}{\color[rgb]{0,0,0}$a_3$}%
}}}}
\put(2296,-1141){\makebox(0,0)[lb]{\smash{{\SetFigFontNFSS{12}{14.4}{\familydefault}{\mddefault}{\updefault}{\color[rgb]{0,0,0}$!$}%
}}}}
\put(2026,-196){\makebox(0,0)[lb]{\smash{{\SetFigFontNFSS{12}{14.4}{\familydefault}{\mddefault}{\updefault}{\color[rgb]{0,0,0}$a_1$}%
}}}}
\put(2296,-151){\makebox(0,0)[lb]{\smash{{\SetFigFontNFSS{12}{14.4}{\familydefault}{\mddefault}{\updefault}{\color[rgb]{0,0,0}$!$}%
}}}}
\put(541,-691){\makebox(0,0)[lb]{\smash{{\SetFigFontNFSS{12}{14.4}{\familydefault}{\mddefault}{\updefault}{\color[rgb]{0,0,0}$a_2$}%
}}}}
\put(811,-646){\makebox(0,0)[lb]{\smash{{\SetFigFontNFSS{12}{14.4}{\familydefault}{\mddefault}{\updefault}{\color[rgb]{0,0,0}$!$}%
}}}}
\put(541,-1186){\makebox(0,0)[lb]{\smash{{\SetFigFontNFSS{12}{14.4}{\familydefault}{\mddefault}{\updefault}{\color[rgb]{0,0,0}$a_3$}%
}}}}
\put(811,-1141){\makebox(0,0)[lb]{\smash{{\SetFigFontNFSS{12}{14.4}{\familydefault}{\mddefault}{\updefault}{\color[rgb]{0,0,0}$!$}%
}}}}
\put(541,-196){\makebox(0,0)[lb]{\smash{{\SetFigFontNFSS{12}{14.4}{\familydefault}{\mddefault}{\updefault}{\color[rgb]{0,0,0}$a_1$}%
}}}}
\put(811,-151){\makebox(0,0)[lb]{\smash{{\SetFigFontNFSS{12}{14.4}{\familydefault}{\mddefault}{\updefault}{\color[rgb]{0,0,0}$!$}%
}}}}
\put(136,-2221){\makebox(0,0)[lb]{\smash{{\SetFigFontNFSS{12}{14.4}{\familydefault}{\mddefault}{\updefault}{\color[rgb]{0,0,0}$O(\max\{p,r\}^2)$}%
}}}}
\put(811,-1771){\makebox(0,0)[lb]{\smash{{\SetFigFontNFSS{12}{14.4}{\familydefault}{\mddefault}{\updefault}{\color[rgb]{0,0,0}$p$}%
}}}}
\put(271,-1546){\makebox(0,0)[lb]{\smash{{\SetFigFontNFSS{12}{14.4}{\familydefault}{\mddefault}{\updefault}{\color[rgb]{0,0,0}$r(p+1)$}%
}}}}
\put(1306,299){\makebox(0,0)[lb]{\smash{{\SetFigFontNFSS{12}{14.4}{\familydefault}{\mddefault}{\updefault}{\color[rgb]{0,0,0}$e$}%
}}}}
\end{picture}%

%% file: weak-squaring-chain.pstex_t
\begin{picture}(0,0)%
\includegraphics{weak-squaring-chain.pstex}%
\end{picture}%
\setlength{\unitlength}{4144sp}%
\begingroup\makeatletter\ifx\SetFigFontNFSS\undefined%
\gdef\SetFigFontNFSS#1#2#3#4#5{%
  \reset@font\fontsize{#1}{#2pt}%
  \fontfamily{#3}\fontseries{#4}\fontshape{#5}%
  \selectfont}%
\fi\endgroup%
\begin{picture}(3672,2911)(166,-2150)
\put(361,479){\makebox(0,0)[lb]{\smash{{\SetFigFontNFSS{12}{14.4}{\familydefault}{\mddefault}{\updefault}{\color[rgb]{0,0,0}$\$$}%
}}}}
\put(451,-16){\makebox(0,0)[lb]{\smash{{\SetFigFontNFSS{12}{14.4}{\familydefault}{\mddefault}{\updefault}{\color[rgb]{0,0,0}$p$}%
}}}}
\put(316,-1771){\makebox(0,0)[lb]{\smash{{\SetFigFontNFSS{12}{14.4}{\familydefault}{\mddefault}{\updefault}{\color[rgb]{0,0,0}$r$}%
}}}}
\put(3106,-16){\makebox(0,0)[lb]{\smash{{\SetFigFontNFSS{12}{14.4}{\familydefault}{\mddefault}{\updefault}{\color[rgb]{0,0,0}$q$}%
}}}}
\put(181,-1996){\makebox(0,0)[lb]{\smash{{\SetFigFontNFSS{12}{14.4}{\familydefault}{\mddefault}{\updefault}{\color[rgb]{0,0,0}$O(\max\{p,q,r,s\})$}%
}}}}
\put(361,-961){\makebox(0,0)[lb]{\smash{{\SetFigFontNFSS{12}{14.4}{\familydefault}{\mddefault}{\updefault}{\color[rgb]{0,0,0}$\$$}%
}}}}
\put(586,-1006){\makebox(0,0)[lb]{\smash{{\SetFigFontNFSS{12}{14.4}{\familydefault}{\mddefault}{\updefault}{\color[rgb]{0,0,0}$c_1$}%
}}}}
\put(1306,-1006){\makebox(0,0)[lb]{\smash{{\SetFigFontNFSS{12}{14.4}{\familydefault}{\mddefault}{\updefault}{\color[rgb]{0,0,0}$a_2$}%
}}}}
\put(1576,-961){\makebox(0,0)[lb]{\smash{{\SetFigFontNFSS{12}{14.4}{\familydefault}{\mddefault}{\updefault}{\color[rgb]{0,0,0}$!$}%
}}}}
\put(901,-1501){\makebox(0,0)[lb]{\smash{{\SetFigFontNFSS{12}{14.4}{\familydefault}{\mddefault}{\updefault}{\color[rgb]{0,0,0}$c_2$}%
}}}}
\put(676,-1456){\makebox(0,0)[lb]{\smash{{\SetFigFontNFSS{12}{14.4}{\familydefault}{\mddefault}{\updefault}{\color[rgb]{0,0,0}$\$$}%
}}}}
\put(1666,-1501){\makebox(0,0)[lb]{\smash{{\SetFigFontNFSS{12}{14.4}{\familydefault}{\mddefault}{\updefault}{\color[rgb]{0,0,0}$a_3$}%
}}}}
\put(1936,-1456){\makebox(0,0)[lb]{\smash{{\SetFigFontNFSS{12}{14.4}{\familydefault}{\mddefault}{\updefault}{\color[rgb]{0,0,0}$!$}%
}}}}
\put(2386,-1456){\makebox(0,0)[lb]{\smash{{\SetFigFontNFSS{12}{14.4}{\familydefault}{\mddefault}{\updefault}{\color[rgb]{0,0,0}$\$$}%
}}}}
\put(2611,-1501){\makebox(0,0)[lb]{\smash{{\SetFigFontNFSS{12}{14.4}{\familydefault}{\mddefault}{\updefault}{\color[rgb]{0,0,0}$d_2$}%
}}}}
\put(3331,-1501){\makebox(0,0)[lb]{\smash{{\SetFigFontNFSS{12}{14.4}{\familydefault}{\mddefault}{\updefault}{\color[rgb]{0,0,0}$b_3$}%
}}}}
\put(3601,-1456){\makebox(0,0)[lb]{\smash{{\SetFigFontNFSS{12}{14.4}{\familydefault}{\mddefault}{\updefault}{\color[rgb]{0,0,0}$!$}%
}}}}
\put(2971,-1006){\makebox(0,0)[lb]{\smash{{\SetFigFontNFSS{12}{14.4}{\familydefault}{\mddefault}{\updefault}{\color[rgb]{0,0,0}$b_2$}%
}}}}
\put(3241,-961){\makebox(0,0)[lb]{\smash{{\SetFigFontNFSS{12}{14.4}{\familydefault}{\mddefault}{\updefault}{\color[rgb]{0,0,0}$!$}%
}}}}
\put(2611,-511){\makebox(0,0)[lb]{\smash{{\SetFigFontNFSS{12}{14.4}{\familydefault}{\mddefault}{\updefault}{\color[rgb]{0,0,0}$b_1$}%
}}}}
\put(2881,-466){\makebox(0,0)[lb]{\smash{{\SetFigFontNFSS{12}{14.4}{\familydefault}{\mddefault}{\updefault}{\color[rgb]{0,0,0}$!$}%
}}}}
\put(946,-511){\makebox(0,0)[lb]{\smash{{\SetFigFontNFSS{12}{14.4}{\familydefault}{\mddefault}{\updefault}{\color[rgb]{0,0,0}$a_1$}%
}}}}
\put(1216,-466){\makebox(0,0)[lb]{\smash{{\SetFigFontNFSS{12}{14.4}{\familydefault}{\mddefault}{\updefault}{\color[rgb]{0,0,0}$!$}%
}}}}
\put(2161,-2086){\makebox(0,0)[lb]{\smash{{\SetFigFontNFSS{12}{14.4}{\familydefault}{\mddefault}{\updefault}{\color[rgb]{1,0,0}$\$ A$}%
}}}}
\put(3601,-1771){\makebox(0,0)[lb]{\smash{{\SetFigFontNFSS{12}{14.4}{\familydefault}{\mddefault}{\updefault}{\color[rgb]{0,0,0}$s$}%
}}}}
\end{picture}%

%% file: weak-squaring-chain-to-squaring.pstex_t
\begin{picture}(0,0)%
\includegraphics{weak-squaring-chain-to-squaring.pstex}%
\end{picture}%
\setlength{\unitlength}{4144sp}%
\begingroup\makeatletter\ifx\SetFigFontNFSS\undefined%
\gdef\SetFigFontNFSS#1#2#3#4#5{%
  \reset@font\fontsize{#1}{#2pt}%
  \fontfamily{#3}\fontseries{#4}\fontshape{#5}%
  \selectfont}%
\fi\endgroup%
\begin{picture}(2679,2821)(124,-2060)
\put(226,485){\makebox(0,0)[lb]{\smash{{\SetFigFontNFSS{12}{14.4}{\familydefault}{\mddefault}{\updefault}{\color[rgb]{0,0,0}$\$$}%
}}}}
\put(271,-1726){\makebox(0,0)[lb]{\smash{{\SetFigFontNFSS{12}{14.4}{\familydefault}{\mddefault}{\updefault}{\color[rgb]{0,0,0}$r$}%
}}}}
\put(226,-16){\makebox(0,0)[lb]{\smash{{\SetFigFontNFSS{12}{14.4}{\familydefault}{\mddefault}{\updefault}{\color[rgb]{0,0,0}$p$}%
}}}}
\put(2251,-16){\makebox(0,0)[lb]{\smash{{\SetFigFontNFSS{12}{14.4}{\familydefault}{\mddefault}{\updefault}{\color[rgb]{0,0,0}$q$}%
}}}}
\put(181,-1951){\makebox(0,0)[lb]{\smash{{\SetFigFontNFSS{12}{14.4}{\familydefault}{\mddefault}{\updefault}{\color[rgb]{0,0,0}$O(\max\{p,q,r,s\})$}%
}}}}
\put(1126,-421){\makebox(0,0)[lb]{\smash{{\SetFigFontNFSS{12}{14.4}{\familydefault}{\mddefault}{\updefault}{\color[rgb]{0,0,0}$!$}%
}}}}
\put(1126,-916){\makebox(0,0)[lb]{\smash{{\SetFigFontNFSS{12}{14.4}{\familydefault}{\mddefault}{\updefault}{\color[rgb]{0,0,0}$!$}%
}}}}
\put(676,-1456){\makebox(0,0)[lb]{\smash{{\SetFigFontNFSS{12}{14.4}{\familydefault}{\mddefault}{\updefault}{\color[rgb]{0,0,0}$a_3$}%
}}}}
\put(946,-1411){\makebox(0,0)[lb]{\smash{{\SetFigFontNFSS{12}{14.4}{\familydefault}{\mddefault}{\updefault}{\color[rgb]{0,0,0}$!$}%
}}}}
\put(2566,-916){\makebox(0,0)[lb]{\smash{{\SetFigFontNFSS{12}{14.4}{\familydefault}{\mddefault}{\updefault}{\color[rgb]{0,0,0}$!$}%
}}}}
\put(1801,-466){\makebox(0,0)[lb]{\smash{{\SetFigFontNFSS{12}{14.4}{\familydefault}{\mddefault}{\updefault}{\color[rgb]{0,0,0}$b_1$}%
}}}}
\put(2071,-421){\makebox(0,0)[lb]{\smash{{\SetFigFontNFSS{12}{14.4}{\familydefault}{\mddefault}{\updefault}{\color[rgb]{0,0,0}$!$}%
}}}}
\put(2116,-1456){\makebox(0,0)[lb]{\smash{{\SetFigFontNFSS{12}{14.4}{\familydefault}{\mddefault}{\updefault}{\color[rgb]{0,0,0}$b_3$}%
}}}}
\put(2386,-1411){\makebox(0,0)[lb]{\smash{{\SetFigFontNFSS{12}{14.4}{\familydefault}{\mddefault}{\updefault}{\color[rgb]{0,0,0}$!$}%
}}}}
\put(1756,-1996){\makebox(0,0)[lb]{\smash{{\SetFigFontNFSS{12}{14.4}{\familydefault}{\mddefault}{\updefault}{\color[rgb]{1,0,0}$\$ A$}%
}}}}
\put(2566,-1726){\makebox(0,0)[lb]{\smash{{\SetFigFontNFSS{12}{14.4}{\familydefault}{\mddefault}{\updefault}{\color[rgb]{0,0,0}$s$}%
}}}}
\put(406,-961){\makebox(0,0)[lb]{\smash{{\SetFigFontNFSS{12}{14.4}{\familydefault}{\mddefault}{\updefault}{\color[rgb]{0,0,0}$a_2\bowtie c_2$}%
}}}}
\put(406,-466){\makebox(0,0)[lb]{\smash{{\SetFigFontNFSS{12}{14.4}{\familydefault}{\mddefault}{\updefault}{\color[rgb]{0,0,0}$a_1\bowtie c_1$}%
}}}}
\put(1846,-961){\makebox(0,0)[lb]{\smash{{\SetFigFontNFSS{12}{14.4}{\familydefault}{\mddefault}{\updefault}{\color[rgb]{0,0,0}$b_2\bowtie d_2$}%
}}}}
\end{picture}%

%% file: weak-squaring-chain-reduced.pstex_t
\begin{picture}(0,0)%
\includegraphics{weak-squaring-chain-reduced.pstex}%
\end{picture}%
\setlength{\unitlength}{4144sp}%
\begingroup\makeatletter\ifx\SetFigFontNFSS\undefined%
\gdef\SetFigFontNFSS#1#2#3#4#5{%
  \reset@font\fontsize{#1}{#2pt}%
  \fontfamily{#3}\fontseries{#4}\fontshape{#5}%
  \selectfont}%
\fi\endgroup%
\begin{picture}(5112,2875)(121,-2204)
\put(541,-1051){\makebox(0,0)[lb]{\smash{{\SetFigFontNFSS{12}{14.4}{\familydefault}{\mddefault}{\updefault}{\color[rgb]{0,0,0}$a_3$}%
}}}}
\put(811,-1006){\makebox(0,0)[lb]{\smash{{\SetFigFontNFSS{12}{14.4}{\familydefault}{\mddefault}{\updefault}{\color[rgb]{0,0,0}$!$}%
}}}}
\put(1441,-1051){\makebox(0,0)[lb]{\smash{{\SetFigFontNFSS{12}{14.4}{\familydefault}{\mddefault}{\updefault}{\color[rgb]{0,0,0}$a_3$}%
}}}}
\put(1711,-1006){\makebox(0,0)[lb]{\smash{{\SetFigFontNFSS{12}{14.4}{\familydefault}{\mddefault}{\updefault}{\color[rgb]{0,0,0}$!$}%
}}}}
\put(2341,-1051){\makebox(0,0)[lb]{\smash{{\SetFigFontNFSS{12}{14.4}{\familydefault}{\mddefault}{\updefault}{\color[rgb]{0,0,0}$a_3$}%
}}}}
\put(2611,-1006){\makebox(0,0)[lb]{\smash{{\SetFigFontNFSS{12}{14.4}{\familydefault}{\mddefault}{\updefault}{\color[rgb]{0,0,0}$!$}%
}}}}
\put(3241,-1051){\makebox(0,0)[lb]{\smash{{\SetFigFontNFSS{12}{14.4}{\familydefault}{\mddefault}{\updefault}{\color[rgb]{0,0,0}$a_3$}%
}}}}
\put(3511,-1006){\makebox(0,0)[lb]{\smash{{\SetFigFontNFSS{12}{14.4}{\familydefault}{\mddefault}{\updefault}{\color[rgb]{0,0,0}$!$}%
}}}}
\put(586,389){\makebox(0,0)[lb]{\smash{{\SetFigFontNFSS{12}{14.4}{\familydefault}{\mddefault}{\updefault}{\color[rgb]{0,0,0}$\$$}%
}}}}
\put(136,-2131){\makebox(0,0)[lb]{\smash{{\SetFigFontNFSS{12}{14.4}{\familydefault}{\mddefault}{\updefault}{\color[rgb]{0,0,0}$O(\max\{p,r\}^2)$}%
}}}}
\put(1036,-1636){\makebox(0,0)[lb]{\smash{{\SetFigFontNFSS{12}{14.4}{\familydefault}{\mddefault}{\updefault}{\color[rgb]{0,0,0}$p$}%
}}}}
\put(181,-1456){\makebox(0,0)[lb]{\smash{{\SetFigFontNFSS{12}{14.4}{\familydefault}{\mddefault}{\updefault}{\color[rgb]{0,0,0}$r(p+1)$}%
}}}}
\put(4816,-61){\makebox(0,0)[lb]{\smash{{\SetFigFontNFSS{12}{14.4}{\familydefault}{\mddefault}{\updefault}{\color[rgb]{0,0,0}$q$}%
}}}}
\put(5131,-961){\makebox(0,0)[lb]{\smash{{\SetFigFontNFSS{12}{14.4}{\familydefault}{\mddefault}{\updefault}{\color[rgb]{0,0,0}$!$}%
}}}}
\put(4366,-511){\makebox(0,0)[lb]{\smash{{\SetFigFontNFSS{12}{14.4}{\familydefault}{\mddefault}{\updefault}{\color[rgb]{0,0,0}$b_1$}%
}}}}
\put(4636,-466){\makebox(0,0)[lb]{\smash{{\SetFigFontNFSS{12}{14.4}{\familydefault}{\mddefault}{\updefault}{\color[rgb]{0,0,0}$!$}%
}}}}
\put(4681,-1501){\makebox(0,0)[lb]{\smash{{\SetFigFontNFSS{12}{14.4}{\familydefault}{\mddefault}{\updefault}{\color[rgb]{0,0,0}$b_3$}%
}}}}
\put(4951,-1456){\makebox(0,0)[lb]{\smash{{\SetFigFontNFSS{12}{14.4}{\familydefault}{\mddefault}{\updefault}{\color[rgb]{0,0,0}$!$}%
}}}}
\put(4321,-2041){\makebox(0,0)[lb]{\smash{{\SetFigFontNFSS{12}{14.4}{\familydefault}{\mddefault}{\updefault}{\color[rgb]{1,0,0}$\$ A$}%
}}}}
\put(991,-556){\makebox(0,0)[lb]{\smash{{\SetFigFontNFSS{12}{14.4}{\familydefault}{\mddefault}{\updefault}{\color[rgb]{0,0,0}$!$}%
}}}}
\put(991,-61){\makebox(0,0)[lb]{\smash{{\SetFigFontNFSS{12}{14.4}{\familydefault}{\mddefault}{\updefault}{\color[rgb]{0,0,0}$!$}%
}}}}
\put(1891,-61){\makebox(0,0)[lb]{\smash{{\SetFigFontNFSS{12}{14.4}{\familydefault}{\mddefault}{\updefault}{\color[rgb]{0,0,0}$!$}%
}}}}
\put(1891,-556){\makebox(0,0)[lb]{\smash{{\SetFigFontNFSS{12}{14.4}{\familydefault}{\mddefault}{\updefault}{\color[rgb]{0,0,0}$!$}%
}}}}
\put(2791,-61){\makebox(0,0)[lb]{\smash{{\SetFigFontNFSS{12}{14.4}{\familydefault}{\mddefault}{\updefault}{\color[rgb]{0,0,0}$!$}%
}}}}
\put(2791,-556){\makebox(0,0)[lb]{\smash{{\SetFigFontNFSS{12}{14.4}{\familydefault}{\mddefault}{\updefault}{\color[rgb]{0,0,0}$!$}%
}}}}
\put(3691,-61){\makebox(0,0)[lb]{\smash{{\SetFigFontNFSS{12}{14.4}{\familydefault}{\mddefault}{\updefault}{\color[rgb]{0,0,0}$!$}%
}}}}
\put(3691,-556){\makebox(0,0)[lb]{\smash{{\SetFigFontNFSS{12}{14.4}{\familydefault}{\mddefault}{\updefault}{\color[rgb]{0,0,0}$!$}%
}}}}
\put(271,-106){\makebox(0,0)[lb]{\smash{{\SetFigFontNFSS{12}{14.4}{\familydefault}{\mddefault}{\updefault}{\color[rgb]{0,0,0}$a_1\bowtie c_1$}%
}}}}
\put(271,-601){\makebox(0,0)[lb]{\smash{{\SetFigFontNFSS{12}{14.4}{\familydefault}{\mddefault}{\updefault}{\color[rgb]{0,0,0}$a_2\bowtie c_2$}%
}}}}
\put(1171,-106){\makebox(0,0)[lb]{\smash{{\SetFigFontNFSS{12}{14.4}{\familydefault}{\mddefault}{\updefault}{\color[rgb]{0,0,0}$a_1\bowtie c_1$}%
}}}}
\put(1171,-601){\makebox(0,0)[lb]{\smash{{\SetFigFontNFSS{12}{14.4}{\familydefault}{\mddefault}{\updefault}{\color[rgb]{0,0,0}$a_2\bowtie c_2$}%
}}}}
\put(2071,-106){\makebox(0,0)[lb]{\smash{{\SetFigFontNFSS{12}{14.4}{\familydefault}{\mddefault}{\updefault}{\color[rgb]{0,0,0}$a_1\bowtie c_1$}%
}}}}
\put(2071,-601){\makebox(0,0)[lb]{\smash{{\SetFigFontNFSS{12}{14.4}{\familydefault}{\mddefault}{\updefault}{\color[rgb]{0,0,0}$a_2\bowtie c_2$}%
}}}}
\put(2971,-106){\makebox(0,0)[lb]{\smash{{\SetFigFontNFSS{12}{14.4}{\familydefault}{\mddefault}{\updefault}{\color[rgb]{0,0,0}$a_1\bowtie c_1$}%
}}}}
\put(2971,-601){\makebox(0,0)[lb]{\smash{{\SetFigFontNFSS{12}{14.4}{\familydefault}{\mddefault}{\updefault}{\color[rgb]{0,0,0}$a_2\bowtie c_2$}%
}}}}
\put(4411,-1006){\makebox(0,0)[lb]{\smash{{\SetFigFontNFSS{12}{14.4}{\familydefault}{\mddefault}{\updefault}{\color[rgb]{0,0,0}$b_2\bowtie d_2$}%
}}}}
\put(5086,-1771){\makebox(0,0)[lb]{\smash{{\SetFigFontNFSS{12}{14.4}{\familydefault}{\mddefault}{\updefault}{\color[rgb]{0,0,0}$s$}%
}}}}
\end{picture}%

%% file: list-of-constants-generation.pstex_t
\begin{picture}(0,0)%
\includegraphics{list-of-constants-generation.pstex}%
\end{picture}%
\setlength{\unitlength}{4144sp}%
\begingroup\makeatletter\ifx\SetFigFontNFSS\undefined%
\gdef\SetFigFontNFSS#1#2#3#4#5{%
  \reset@font\fontsize{#1}{#2pt}%
  \fontfamily{#3}\fontseries{#4}\fontshape{#5}%
  \selectfont}%
\fi\endgroup%
\begin{picture}(7185,7555)(211,-6704)
\put(1306,614){\makebox(0,0)[lb]{\smash{{\SetFigFontNFSS{12}{14.4}{\familydefault}{\mddefault}{\updefault}{\color[rgb]{0,0,0}$\$\UIntT\liv\UIntT\li\ListT\UIntT$}%
}}}}
\put(1306,-16){\makebox(0,0)[lb]{\smash{{\SetFigFontNFSS{12}{14.4}{\familydefault}{\mddefault}{\updefault}{\color[rgb]{0,0,0}$\UIntT\li\ListT\UIntT$}%
}}}}
\put(1126,344){\makebox(0,0)[lb]{\smash{{\SetFigFontNFSS{12}{14.4}{\familydefault}{\mddefault}{\updefault}{\color[rgb]{0,0,0}\scriptsize$\liv\!I$\normalsize}%
}}}}
\put(1126,-286){\makebox(0,0)[lb]{\smash{{\SetFigFontNFSS{12}{14.4}{\familydefault}{\mddefault}{\updefault}{\color[rgb]{0,0,0}\scriptsize$\li\!I$\normalsize}%
}}}}
\put(1171,-916){\makebox(0,0)[lb]{\smash{{\SetFigFontNFSS{12}{14.4}{\familydefault}{\mddefault}{\updefault}{\color[rgb]{0,0,0}\scriptsize$\forall I$\normalsize}%
}}}}
\put(1126,-1546){\makebox(0,0)[lb]{\smash{{\SetFigFontNFSS{12}{14.4}{\familydefault}{\mddefault}{\updefault}{\color[rgb]{0,0,0}\scriptsize$\li\!I$\normalsize}%
}}}}
\put(1081,-2176){\makebox(0,0)[lb]{\smash{{\SetFigFontNFSS{12}{14.4}{\familydefault}{\mddefault}{\updefault}{\color[rgb]{0,0,0}\scriptsize$\li\!E$\normalsize}%
}}}}
\put(1081,-4066){\makebox(0,0)[lb]{\smash{{\SetFigFontNFSS{12}{14.4}{\familydefault}{\mddefault}{\updefault}{\color[rgb]{0,0,0}\scriptsize$\li\!E$\normalsize}%
}}}}
\put(1081,-4696){\makebox(0,0)[lb]{\smash{{\SetFigFontNFSS{12}{14.4}{\familydefault}{\mddefault}{\updefault}{\color[rgb]{0,0,0}\scriptsize$\li\!E$\normalsize}%
}}}}
\put(1711,-4471){\makebox(0,0)[lb]{\smash{{\SetFigFontNFSS{12}{14.4}{\familydefault}{\mddefault}{\updefault}{\color[rgb]{0,0,0}$I$}%
}}}}
\put(1306,-646){\makebox(0,0)[lb]{\smash{{\SetFigFontNFSS{12}{14.4}{\familydefault}{\mddefault}{\updefault}{\color[rgb]{0,0,0}$\ListT\,\UIntT\equiv \forall \alpha.!(\UIntT\li\alpha\li\alpha)\li\$(\alpha\li\alpha)$}%
}}}}
\put(1306,-1276){\makebox(0,0)[lb]{\smash{{\SetFigFontNFSS{12}{14.4}{\familydefault}{\mddefault}{\updefault}{\color[rgb]{0,0,0}$!(\UIntT\li\alpha\li\alpha)\li\$(\alpha\li\alpha)$}%
}}}}
\put(1486,-2041){\makebox(0,0)[lb]{\smash{{\SetFigFontNFSS{12}{14.4}{\familydefault}{\mddefault}{\updefault}{\color[rgb]{0,0,0}$\$(((\delta\li\delta)\li\alpha)\li(\delta\li\delta)\li\alpha)$}%
}}}}
\put(1261,-3121){\makebox(0,0)[lb]{\smash{{\SetFigFontNFSS{12}{14.4}{\familydefault}{\mddefault}{\updefault}{\color[rgb]{0,0,0}$\$(\alpha\li\alpha)$}%
}}}}
\put(1126,-2806){\makebox(0,0)[lb]{\smash{{\SetFigFontNFSS{12}{14.4}{\familydefault}{\mddefault}{\updefault}{\color[rgb]{0,0,0}\scriptsize$\li\!I$\normalsize}%
}}}}
\put(1126,-3436){\makebox(0,0)[lb]{\smash{{\SetFigFontNFSS{12}{14.4}{\familydefault}{\mddefault}{\updefault}{\color[rgb]{0,0,0}\scriptsize$\li\!I$\normalsize}%
}}}}
\put(3826,-3301){\makebox(0,0)[lb]{\smash{{\SetFigFontNFSS{12}{14.4}{\familydefault}{\mddefault}{\updefault}{\color[rgb]{0,0,0}$\$$}%
}}}}
\put(1666,-5011){\makebox(0,0)[lb]{\smash{{\SetFigFontNFSS{12}{14.4}{\familydefault}{\mddefault}{\updefault}{\color[rgb]{0,0,0}\scriptsize$\li\!I$\normalsize}%
}}}}
\put(1801,-4741){\makebox(0,0)[lb]{\smash{{\SetFigFontNFSS{12}{14.4}{\familydefault}{\mddefault}{\updefault}{\color[rgb]{0,0,0}$(\delta\li\delta)\li\alpha$}%
}}}}
\put(1261,-5641){\makebox(0,0)[lb]{\smash{{\SetFigFontNFSS{12}{14.4}{\familydefault}{\mddefault}{\updefault}{\color[rgb]{0,0,0}$((\delta\li\delta)\li\alpha)\li(\delta\li\delta)\li\alpha$}%
}}}}
\put(1261,-5641){\makebox(0,0)[lb]{\smash{{\SetFigFontNFSS{12}{14.4}{\familydefault}{\mddefault}{\updefault}{\color[rgb]{0,0,0}$((\delta\li\delta)\li\alpha)\li(\delta\li\delta)\li\alpha$}%
}}}}
\put(1801,-5326){\makebox(0,0)[lb]{\smash{{\SetFigFontNFSS{12}{14.4}{\familydefault}{\mddefault}{\updefault}{\color[rgb]{0,0,0}$\alpha$}%
}}}}
\put(1261,-3796){\makebox(0,0)[lb]{\smash{{\SetFigFontNFSS{12}{14.4}{\familydefault}{\mddefault}{\updefault}{\color[rgb]{0,0,0}$\alpha$}%
}}}}
\put(1801,-4201){\makebox(0,0)[lb]{\smash{{\SetFigFontNFSS{12}{14.4}{\familydefault}{\mddefault}{\updefault}{\color[rgb]{0,0,0}$\delta\li\delta$}%
}}}}
\put(4276,-2581){\makebox(0,0)[lb]{\smash{{\SetFigFontNFSS{12}{14.4}{\familydefault}{\mddefault}{\updefault}{\color[rgb]{0,0,0}\scriptsize$\li\!E$\normalsize}%
}}}}
\put(5401,-3166){\makebox(0,0)[lb]{\smash{{\SetFigFontNFSS{12}{14.4}{\familydefault}{\mddefault}{\updefault}{\color[rgb]{0,0,0}\scriptsize$\li\!I$\normalsize}%
}}}}
\put(5401,-3751){\makebox(0,0)[lb]{\smash{{\SetFigFontNFSS{12}{14.4}{\familydefault}{\mddefault}{\updefault}{\color[rgb]{0,0,0}\scriptsize$\li\!I$\normalsize}%
}}}}
\put(5356,-4381){\makebox(0,0)[lb]{\smash{{\SetFigFontNFSS{12}{14.4}{\familydefault}{\mddefault}{\updefault}{\color[rgb]{0,0,0}\scriptsize$\li\!E$\normalsize}%
}}}}
\put(6031,-4651){\makebox(0,0)[lb]{\smash{{\SetFigFontNFSS{12}{14.4}{\familydefault}{\mddefault}{\updefault}{\color[rgb]{0,0,0}\scriptsize$\li\!E$\normalsize}%
}}}}
\put(4591,-2446){\makebox(0,0)[lb]{\smash{{\SetFigFontNFSS{12}{14.4}{\familydefault}{\mddefault}{\updefault}{\color[rgb]{0,0,0}$!(((\delta\li\delta)\li\alpha)\li(\delta\li\delta)\li\alpha)$}%
}}}}
\put(5356,-5011){\makebox(0,0)[lb]{\smash{{\SetFigFontNFSS{12}{14.4}{\familydefault}{\mddefault}{\updefault}{\color[rgb]{0,0,0}\scriptsize$\li\!E$\normalsize}%
}}}}
\put(6571,-4966){\makebox(0,0)[lb]{\smash{{\SetFigFontNFSS{12}{14.4}{\familydefault}{\mddefault}{\updefault}{\color[rgb]{0,0,0}$I$}%
}}}}
\put(5536,-3481){\makebox(0,0)[lb]{\smash{{\SetFigFontNFSS{12}{14.4}{\familydefault}{\mddefault}{\updefault}{\color[rgb]{0,0,0}$(\delta\li\delta)\li\alpha$}%
}}}}
\put(4321,-5551){\makebox(0,0)[lb]{\smash{{\SetFigFontNFSS{12}{14.4}{\familydefault}{\mddefault}{\updefault}{\color[rgb]{0,0,0}\scriptsize$\forall E$\normalsize}%
}}}}
\put(5536,-6226){\makebox(0,0)[lb]{\smash{{\SetFigFontNFSS{12}{14.4}{\familydefault}{\mddefault}{\updefault}{\color[rgb]{0,0,0}$\tc{blue}{!(\UIntT\li\alpha\li\alpha)}$}%
}}}}
\put(5896,-5821){\makebox(0,0)[lb]{\smash{{\SetFigFontNFSS{12}{14.4}{\familydefault}{\mddefault}{\updefault}{\color[rgb]{0,0,0}$\tc{red}{\$\UIntT}$}%
}}}}
\put(7381,-2941){\makebox(0,0)[lb]{\smash{{\SetFigFontNFSS{12}{14.4}{\familydefault}{\mddefault}{\updefault}{\color[rgb]{0,0,0}$!$}%
}}}}
\put(6211,-5191){\makebox(0,0)[lb]{\smash{{\SetFigFontNFSS{12}{14.4}{\familydefault}{\mddefault}{\updefault}{\color[rgb]{0,0,0}$(\delta\li\delta)\li\alpha$}%
}}}}
\put(5761,-4291){\makebox(0,0)[lb]{\smash{{\SetFigFontNFSS{12}{14.4}{\familydefault}{\mddefault}{\updefault}{\color[rgb]{0,0,0}$\alpha$}%
}}}}
\put(5536,-4066){\makebox(0,0)[lb]{\smash{{\SetFigFontNFSS{12}{14.4}{\familydefault}{\mddefault}{\updefault}{\color[rgb]{0,0,0}$\alpha$}%
}}}}
\put(226,-6631){\makebox(0,0)[lb]{\smash{{\SetFigFontNFSS{12}{14.4}{\familydefault}{\mddefault}{\updefault}{\color[rgb]{0,0,0}$\UIntT\equiv\forall \alpha.!(\alpha\li\alpha)\li\$(\alpha\li\alpha)$}%
}}}}
\end{picture}%

%% file: square-composition-block-diagram.pstex_t
\begin{picture}(0,0)%
\includegraphics{square-composition-block-diagram.pstex}%
\end{picture}%
\setlength{\unitlength}{4144sp}%
\begingroup\makeatletter\ifx\SetFigFontNFSS\undefined%
\gdef\SetFigFontNFSS#1#2#3#4#5{%
  \reset@font\fontsize{#1}{#2pt}%
  \fontfamily{#3}\fontseries{#4}\fontshape{#5}%
  \selectfont}%
\fi\endgroup%
\begin{picture}(5739,6549)(484,-5788)
\put(2431,-736){\makebox(0,0)[lb]{\smash{{\SetFigFontNFSS{12}{14.4}{\familydefault}{\mddefault}{\updefault}{\color[rgb]{0,0,0}$H_1$}%
}}}}
\put(991,-5551){\makebox(0,0)[lb]{\smash{{\SetFigFontNFSS{12}{14.4}{\familydefault}{\mddefault}{\updefault}{\color[rgb]{0,0,0}$\nabla^1_4$}%
}}}}
\put(5311,-4876){\makebox(0,0)[lb]{\smash{{\SetFigFontNFSS{12}{14.4}{\familydefault}{\mddefault}{\updefault}{\color[rgb]{0,0,0}$\rmshare{1}{2}{3}{m}$}%
}}}}
\put(5311,-3301){\makebox(0,0)[lb]{\smash{{\SetFigFontNFSS{12}{14.4}{\familydefault}{\mddefault}{\updefault}{\color[rgb]{0,0,0}$\rmshare{1}{6}{3}{m}$}%
}}}}
\put(5311,-4111){\makebox(0,0)[lb]{\smash{{\SetFigFontNFSS{12}{14.4}{\familydefault}{\mddefault}{\updefault}{\color[rgb]{0,0,0}$\rmshare{1}{4}{3}{m}$}%
}}}}
\put(2746,-2986){\makebox(0,0)[lb]{\smash{{\SetFigFontNFSS{12}{14.4}{\familydefault}{\mddefault}{\updefault}{\color[rgb]{0,0,0}$\mshsqcomp{1}{3}{3}{m}$}%
}}}}
\put(5851,164){\makebox(0,0)[lb]{\smash{{\SetFigFontNFSS{12}{14.4}{\familydefault}{\mddefault}{\updefault}{\color[rgb]{0,0,0}$\sqcomp{1}{3}{1}$}%
}}}}
\put(2611,-16){\makebox(0,0)[lb]{\smash{{\SetFigFontNFSS{12}{14.4}{\familydefault}{\mddefault}{\updefault}{\color[rgb]{0,0,0}$F$}%
}}}}
\put(3736,-736){\makebox(0,0)[lb]{\smash{{\SetFigFontNFSS{12}{14.4}{\familydefault}{\mddefault}{\updefault}{\color[rgb]{0,0,0}$H_2$}%
}}}}
\put(5491,-2311){\makebox(0,0)[lb]{\smash{{\SetFigFontNFSS{12}{14.4}{\familydefault}{\mddefault}{\updefault}{\color[rgb]{0,0,0}$\chi$}%
}}}}
\put(4996,-781){\makebox(0,0)[lb]{\smash{{\SetFigFontNFSS{12}{14.4}{\familydefault}{\mddefault}{\updefault}{\color[rgb]{0,0,0}$\overline{\overline{0}}$}%
}}}}
\put(901,-736){\makebox(0,0)[lb]{\smash{{\SetFigFontNFSS{14}{16.8}{\familydefault}{\mddefault}{\updefault}{\color[rgb]{0,0,0}$G$}%
}}}}
\end{picture}%

%% file: overview-WALT.tex
\section{Overviewing \WALT\ technically}
\label{section:Overviewing WALT technically}
Here we recall the main aspects of Weak Affine Light Typing (\WALT), as developed in \cite{Roversi:2007-WALT-FULL}, and which are required to present our results.
Recall that \WALT\ is a type assignment for $\lambda$-terms.
\paragraph{The $\lambda$-terms.}
The $\lambda$-terms are generated by the grammar
$M::=x\mid (\bs x.M)\mid (MM)$, where $x$ belongs to a countable set of $\lambda$-variables.
An abstraction $\bs x.M$ binds the (free) occurrence of $x$ in $M$. Given a term $M$, the set of its free variables, those ones which are not bound, is $\FV{M}$.
A \textit{closed term} has no free variables.
The \textit{cardinality of a free variable in a term}
is $\nocc{x}{M}$ and counts the number of \emph{free occurrences of $x$ in $M$}:
\small
\begin{align*}
\nocc{x}{x}&=1 &
\nocc{x}{y}&=0 &(x \not\equiv y)\\
\nocc{x}{\bs x.M}&=0 &
\nocc{x}{\bs y.M}&=\nocc{x}{M} &(x \not\equiv y)\\
\nocc{x}{MN} &=\nocc{x}{M}+\nocc{x}{N}
\end{align*}
\normalsize
$M\{^{N_1}/_{x_1} \cdots {}^{N_m}/_{x_m}\}$ denotes the usual capture free simultaneous substitution of every $N_i$ for the corresponding $x_i$, with $1\leq i \leq m$.
Parentheses are left-associative, so $((\cdots((MM)M)\cdots)M)$
shortens to $MMM\cdots M$.
A sequence of abstractions
$(\bs x_1.\ldots(\bs x_m.M)\ldots)$ is shortened by
$\bs x_1 \ldots x_m.M$, for any $m$.
\PTV\ is the set of the $\lambda$-terms which are \textit{values}, generated by
$V ::=  x
  \mid (\bs x.M)$, where $M$ is any $\lambda$-term.
The \textit{size of a term}
$\size{M}$ gives the \emph{dimension}  of $M$ as expected:
$\size{x}= 1,
\size{\bs x.M}=\size{M}+1,
\size{MN}=\size{M}+\size{N}+1$.
\paragraph{The types of \WALT.}
They are formulae that belong to the language, generated by the following grammar:
\small
\begin{align*}
A &::= L
        \mid !A
        \mid \$A
\\
L &::=  \alpha
        \mid A\li A
        \mid \$A\liv A
        \mid \forall \alpha.L
\end{align*}
\normalsize
$A$ is the start symbol.
A \textit{modal} formula has form $!A$ or $\$A$, and, in particular, $!A$ is $!$-modal, while $\$A$ is $\$$-modal.
$L$ generates \textit{linear}, or \textit{non modal}, formulae, which are ranged over by $L, L'$. Generic formulae by $A, B, C$, instead.
Notice that the substitution of $\alpha$ in $\forall\alpha.L$ for a linear type $L'$, produces $L\subs{L'}{\alpha}$ which is still linear. Somewhat conversely, a universal quantification cannot hide a modal type.
\input{figure-of-the-type-assignment}
\paragraph{The rules of \WALT.}
Figure~\ref{figure:Weak LAL as a type assignment} gives the deductive rules of \WALT, which deduce judgments
$\Gamma; \Delta; {\mathcal E} \vdash \ta{M}{A}$.
$\Gamma$ and $\Delta$ are sets of type assignments, namely of pairs $\ta{x}{A}$.
${\mathcal E}$ is a set of pairs $(\Theta;\Phi)$ such that both $\Theta$ and $\Phi$ are sets of type assignments as well.
\par
$\dom{\{\ta{x_1}{A_1},\ldots,\ta{x_n}{A_n}\}}
=\{x_1,\ldots,x_n\}$ is the \textit{domain} of any set $\{\ta{x_1}{A_1},\ldots,\ta{x_n}{A_n}\}$ of type assignments.
$\Gamma$ will denote a set of \emph{linear type assignments} $\ta{x}{L}$. Every $x\in\dom{\Gamma}$ is called \emph{linear variable}.
$\Delta$ will denote a set of \emph{linear partially discharged} type assignments. Every $x\in \dom{\Delta}$ is called \emph{linear partially discharged}.
${\mathcal E}$ will denote a set of \emph{partially discharged contexts}.
$\mathcal E$ can be either empty or it contains pairs $(\Theta_1;\Phi_1),\ldots,(\Theta_n;\Phi_n)$ where, for every $i\in\{1,\ldots,n\}$, the following four points hold:
(i) $\Theta_i$ is a set of \emph{elementary partially discharged} type  assignment. Every $x\in \dom{\Theta_i}$ is called \emph{elementary};
(ii) $\Phi_i$ is either empty or it is a singleton $\ta{x}{A}$. We call $x$ \emph{polynomial};
(iii) only one between $\Phi_1,\ldots,\Phi_n$ can be $\emptyset$;
(iv) the domains of any two $\Phi_i$ and $\Phi_j$ are distinct, whenever $i\neq j$.
\par
For every
${\mathcal E}=\bigcup^{n}_{i=1} \{(\Theta_i;\Phi_i)\}$,
$\dom{\mathcal E}$ is
$(\bigcup^{n}_{i=1} \dom{\Theta_i})\cup
        (\bigcup^{n}_{i=1} \dom{\Phi_i})$.
In every of the rules of \WALT\ the domain of two sets of type assignments $\Phi_M$ and $\Phi_N$ may intersect when $\Phi_M$ and $\Phi_N$ are part of two partially discharged contexts ${\mathcal E}_M$ and ${\mathcal E}_N$ that belong to two distinct premises of a rule.
This observation justifies the definition of
${\mathcal E}_M\sqcup{\mathcal E}_N$
that merges ${\mathcal E}_M$ and ${\mathcal E}_N$, preserving the structure of a partially discharged context:
\small
\begin{align*}
&
{\mathcal E}_M
\sqcup
{\mathcal E}_N
=
\\
&\quad
\{(\Theta_M,\Theta_N;\Phi) \mid
        (\Theta_M;\Phi) \in {\mathcal E}_M
        \text{ and }
        (\Theta_N;\Phi) \in {\mathcal E}_N
\}\cup\\
&\quad
\{(\Theta_M;\Phi_M) \mid
        (\Theta_M;\Phi_M) \in {\mathcal E}_M
        \text{ and there is no }
        \Theta_N \text{ such that }
        (\Theta_N;\Phi_M) \text{ in } {\mathcal E}_N
\}
\cup\\
&\quad
\{(\Theta_N;\Phi_N) \mid
        (\Theta_N;\Phi_N) \in {\mathcal E}_N
        \text{ and there is no }
        \Theta_M \text{ such that }
        (\Theta_M;\Phi_N) \text{ in } {\mathcal E}_M
\}
\end{align*}
\normalsize
The sequence ${\mathcal E},(\Theta;\Phi)$ denotes that
$(\Theta;\Phi)\not\in{\mathcal E}$. Also,
${\mathcal E} \sqcup \{(\emptyset;\emptyset)\}
= {\mathcal E} \sqcup \emptyset =  {\mathcal E}$.
In every other cases, the domain of two sets of type assignments that belong to two distinct premises of a rule of \WALT\ \textit{must} be disjoint.
\PTT\ is the subset of \textit{typeable} elements $M$ of \PT, namely,
those for which a deduction $\Pi$ with conclusion
$\Gamma;\Delta;{\mathcal E}\vdash\ta{M}{A}$ exists, denoted by
$\Pi\rhd\Gamma;\Delta;{\mathcal E}\vdash\ta{M}{A}$.

\paragraph{\WALT\ and \SF.}
\WALT\ is a subsystem of \SF. This means that if
$\Gamma;\Delta;{\mathcal E}\vdash\ta{M}{A}$ then $M$ has type $t(A)$ from the set of assumptions
$T(\Gamma;\Delta;{\mathcal E})$ in \SF, where:
\begin{center}
   \begin{tabular}{rcccl}
    \begin{minipage}{2cm}
      {\begin{align*}
        t(\alpha)&=\alpha\\
        t(\forall \alpha. A)
                 &=\forall \alpha.t(A)
      \end{align*}}
    \end{minipage}
   &&
    \begin{minipage}{2cm}
      {\begin{align*}
        t(A\li B)&= t(A)\rightarrow t(B)\\
        t(A\liv B)&= t(A)\rightarrow t(B)
      \end{align*}}
    \end{minipage}
   &&
    \begin{minipage}{2cm}
      {\begin{align*}
        t(\$ A)&=t(A)\\
        t(! A)&=t(A)
      \end{align*}}
    \end{minipage}
   \end{tabular}
\end{center}
and $T$ is the obvious extension of the map $t$ to the types in $\Gamma;\Delta;{\mathcal E}$.
\subsection{On the rules of \WALT}
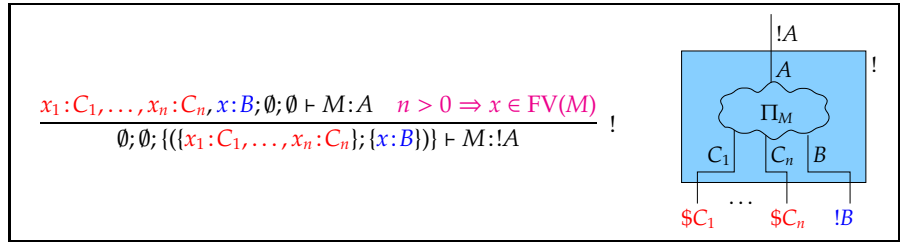
\begin{figure}[ht]
\begin{center}
\fbox{
\begin{tabular}{ccc}
\begin{minipage}[c]{2.5cm}
 \small
 \infer[!]
  {\emptyset;
   \emptyset;
   \{(\{\tc{red}{\ta{x_1}{C_1},\ldots,\ta{x_n}{C_n}}\};
        \{\tc{blue}{\ta{x}{B}}\})\}
   \vdash \ta{M}{\ !A}
  }
  {
   \tc{red}{\ta{x_1}{C_1},\ldots,\ta{x_n}{C_n}},
   \tc{blue}{\ta{x}{B}};
   \emptyset;
   \emptyset
   \vdash \ta{M}{A}
   &
   \tc{magenta}{n>0 \Rightarrow
   x\in\FV{M}}
  }
  \normalsize
\end{minipage}
&\quad&
\begin{minipage}[c]{2.5cm}
\scalebox{.7}{\input{canonical-bang-box.pstex_t}}
\end{minipage}
\end{tabular}
}
\end{center}
\caption{The canonical instance of $!$-box/rule in \WALT}
\label{figure:The canonical instance of a !-box in WALT}
\end{figure}

The bound on the number of normalization steps of any deduction of \WALT\ is a consequence of the stratified nature that \WALT\ inherits from \ILAL.
``Stratification'' means that every deduction $\Pi$ of \WALT\ can be thought of as it was organized into levels, so that the logical rules of $\Pi$ may be at different depths.
The normalization preserves the levels: if an instance of a rule $R$ in $\Pi$ is at depth $d$, then it will keep being at depth $d$ after any number of normalization steps that, of course, do not erase it.
The only duplication allowed is of deductions $\Pi$ that have undergone an instance $r$ of the $!$ rule, namely the conclusion of $\Pi$ has a $!$-modal type, introduced by $r$. Ideally, the $!$ rule defines a, so called, $!$-box around the deduction that proves its premise. Figure~\ref{figure:The canonical instance of a !-box in WALT} shows, side by side, a canonical instance of the rule $!$ and the $!$-box that would correspond to it if we imagined to associate a proof net notation to the derivations of \WALT.
The hypothesis is that $\Pi_M$, with conclusion of type $A$ and assumptions $C_1,\ldots,C_n,B$, corresponds to the term $M$ that has type $B$ from the set of linear type assignments $\ta{x_1}{C_1},\ldots,\ta{x_n}{C_n},\ta{x}{B}$. The application of the rule $!$ corresponds to putting the $!$-box around $\Pi_M$.
\par
The condition $n>0 \Rightarrow x\in\FV{M}$ assures that the substitution of some closed term $N$, with type $!B$, for $x$ in $M$, cannot yield $M\subs{N}{x}$ that only depends on a single assumption. Namely, we want to avoid that a sequence of normalization steps, can yield a judgment $\emptyset;\emptyset;\{(\{\ta{x_1}{C_1}\};\emptyset)\}\vdash\ta{M'}{\,!A}$, where $!A$ says that it can be duplicated, but whose free assumption says that it cannot, since \WALT\ does not have the contraction on $\$$-modal assumptions.
\par
We remark that the $!$-box can be put around a derivation $\Pi$ that may depend on more than one assumption, letting \WALT\ be a strict generalization of \ILAL, whose $!$-boxes, in the context of \WALT, take the form:
\small
$$\infer[]
 {\emptyset;
  \emptyset;
  \{(\emptyset;\Phi)\}
  \vdash \ta{M}{\ !B}
 }
 {
  \Phi;
  \emptyset;
  \emptyset
  \vdash \ta{M}{B}
  &
  \Phi\subseteq\{\ta{x}{A}\}
 }
$$
\normalsize
The elementary partially discharged assumptions the generalized $!$-boxes may depend on can only be replaced, in the course of the normalization steps, by the conclusion of $\$$-boxes exclusively depending on elementary partially discharged assumptions as well.
\begin{figure}[ht]
\begin{center}
\fbox{
\begin{tabular}{ccc}
\begin{minipage}{5.5cm}
\small
\[\infer[\liv E]
  {
   \Gamma;
   \Delta;
   {\mathcal E}_M
   \sqcup
   {\mathcal E}_N
   \vdash \ta{MN}{D}
  }
  {\begin{array}{l}
    \Gamma;
    \Delta;
    {\mathcal E}_M
    \vdash \ta{M}{\$C\liv A}
    \\
    \emptyset;
    \emptyset;
    {\mathcal E}_N
    \vdash \ta{N}{\$C}
    \\
    {\mathcal E}_N\subseteq\{(\tc{red}{\Theta};\tc{blue}{\emptyset})\}
   \end{array}
   }
\]
\normalsize
\end{minipage}
&\quad&
\begin{minipage}{2.5cm}
\scalebox{.6}{\input{eager-arrow-elim.pstex_t}}
\end{minipage}
\end{tabular}
}
\end{center}
\caption{``Net'' meaning of the rule $\liv E$ in \WALT}
\label{figure:Net meaning of the eager arrow elimination}
\end{figure}
Figure~\ref{figure:Net meaning of the eager arrow elimination} shows, with the help of a net, that such a behavior is a consequence of restricting ${\mathcal E}_N$, in the rule $\liv E$, to the form $\{(\Theta;\emptyset)\}$.
The rule $\liv E$ comes with $\liv I$ that can discharge elementary partially discharged assumptions only when the corresponding polynomial assumption has already been discharged by $\li I_!$, as in Figure~\ref{figure:Net meaning of the eager arrow intro and of the arrow bang intro}. The net in such a figure shows the mandatory discharging order.
\begin{figure}
\begin{center}
\fbox{
\begin{tabular}{rp{.5cm}l}
\begin{minipage}{2.5cm}
\scalebox{.6}{\input{eager-and-modal-arrow.pstex_t}}
\end{minipage}
&&
\begin{minipage}{6cm}
\small
\[ \infer[\li I_!]
  {
   \Gamma;
   \Delta;
   {\mathcal E}\sqcup
   \{(\tc{red}{\Theta};\tc{blue}{\emptyset})\}
   \vdash \ta{\bs x.M}{\ \tc{blue}{!B}\li A}
  }
  {
   \Gamma; \Delta;
   {\mathcal E},
   (\tc{red}{\Theta};\{\tc{blue}{\ta{x}{B}}\})
   \vdash \ta{M}{A}
  }
\]
\[
 \infer[\liv I]
  {
   \Gamma;\Delta;{\mathcal E},
   (\tc{red}{\Theta};\tc{blue}{\emptyset})\vdash \ta{\bs x.M}{\tc{red}{\$C}\liv A}
  }
  {
   \Gamma;\Delta;
   {\mathcal E},(\{\tc{red}{\Theta},\tc{red}{\ta{y}{C}}\}; \tc{blue}{\emptyset})
   \vdash \ta{M}{A}
  }
\]
\normalsize
\end{minipage}
\end{tabular}
}
\end{center}
\caption{``Net'' meaning of the rules $\liv I, \li I_!$ in \WALT}
\label{figure:Net meaning of the eager arrow intro and of the arrow bang intro}
\end{figure}
Finally, Figure~\ref{figure:Net meaning of the arrow bang elim} shows, with the help of a net, how $\li E_!$ consistently forces the application of some given $M$ of type $!B\li C$ to an $N$, of type $!B$, according to an order which reverses the one we must use to apply $\li I_!$, and $\liv I$.
\begin{figure}
\begin{center}
\fbox{
\begin{tabular}{rp{.8cm}l}
\begin{minipage}{5.5cm}
\small
\[
\infer[\li E_!]
  {\Gamma_M, \Gamma_N;
   \Delta_M, \Delta_N;
   {\mathcal E}_M
   \sqcup
   {\mathcal E}_N
   \vdash \ta{MN}{C}
  }
  {\begin{array}{l}
   \Gamma_M;
   \Delta_M;
   {\mathcal E}_M
   \vdash \ta{M}{\,\tc{blue}{!B}\li C}
   \\
   \Gamma_N;
   \Delta_N;
   {\mathcal E}_N
   \vdash \ta{N}{\,\tc{blue}{!B}}
   \\
   {\mathcal E}_M\subseteq
   \{(\tc{red}{\emptyset};\tc{blue}{\Phi_1}),
     \ldots,
     (\tc{red}{\emptyset};\tc{blue}{\Phi_n})\}
   \end{array}
   }
\]
\normalsize
\end{minipage}
&&
\begin{minipage}{3.5cm}
\scalebox{.6}{\input{modal-arrow-elimination.pstex_t}}
\end{minipage}
\end{tabular}
}
\end{center}
\caption{``Net'' meaning of the rule $\li E_!$ in \WALT}
\label{figure:Net meaning of the arrow bang elim}
\end{figure}
\par
Summing up, \WALT\ allows to type $\lambda$-terms more liberally than \ILAL, while keeping the same normalization principles: the stratification is never canceled, and only deductions that, eventually, depend on at most one free variable may be effectively duplicated as effect of the normalization.
This is why \WALT\ does not enjoy a full normalizing procedure, the analogous of the cut elimination for a corresponding sequent calculus formulation, as the coming section recalls.
\subsection{The dynamics of \WALT.}
Recall that the \textit{call-by-name}, or \textit{lazy}, $\beta$-reduction on the $\lambda$-terms is the contextual closure of rewriting relation $(\bs x.M)N\rightarrow_{\text{n}} M\subs{N}{x}$.
The \textit{call-by-value}, or \textit{eager}, $\beta$-reduction, instead, is the contextual closure of $(\bs x.M)N\rightarrow_{\text{v}} M\subs{N}{x}$, where $N\in$\PTV.
\par
The subject reduction of the rules in Figure~\ref{figure:Weak LAL as a type assignment} holds only on the following restriction $\rew$ of $\rightarrow_{\text{n}}\cup\rightarrow_{\text{v}}$:
\par
\begin{align*}
(\bs x.M)N\rew M & \text{ if } \nocc{x}{M}=0\\
(\bs x.M)N\rew M\subs{N}{x} & \text{ if } \nocc{x}{M}=1, \text{ and } N\in\text{\PTV}\\
(\bs x.M)N\rew M\subs{N}{x} & \text{ if } \nocc{x}{M}>1,
                                          \text{ and }
                                          N\in\text{\PTV},
                                          \text{ and }
                                          \FV{N}\subseteq\{y\}
\end{align*}
$\rew^+$ is the transitive closure of $\rew$, while
$\rew^*$ is the reflexive and transitive closure of $\rew$.
$M$ is in $\rew$-normal form, and we write $\nf{M}$, if $\rew$ cannot rewrite $M$ anymore.
\par
We conclude by recalling two main features of \WALT:
\begin{theorem}[Subject reduction (\cite{Roversi:2007-WALT-FULL}).]
$\Gamma;\Delta;{\mathcal E}\vdash\ta{M}{A}$ and $M\rew^*N$, imply
$\Gamma;\Delta;{\mathcal E}\vdash\ta{N}{A}$.
\end{theorem}
\begin{theorem}[Polytime soundness (\cite{Roversi:2007-WALT-FULL}).]
Let $\Pi$ be a derivation of \WALT\ whose conclusion be
$\Gamma;\Delta;{\mathcal E}\vdash\ta{M}{A}$. Let $\dpth{\Pi}$ be the maximal depth of $\Pi$, namely the maximal number of instances of the rules $\$, !$ that we can traverse moving from the conclusion of $\Pi$, to every of its axioms instances.
Then, $M$ normalizes to its normal form $\nf{M}$ in a number of steps, hence in a time, which is $O(\size{\Pi}^{k^{\dpth{\Pi}}})$, for some $k$.
\end{theorem}
\subsection{The combinators of \WALT\ we need to recall}
\label{subsection:Basic combinators}
We recall the main aspects of combinators that can be typed in \WALT, and which are required to show the completeness of \WALT\ w.r.t. \SRN. The details are in \cite{Roversi:2007-WALT-FULL}.
\paragraph{(Binary) Words.}
They are the terms:
\begin{align*}
\BNum{0}&\equiv\bs 01y.y
\\
\BNum{2^m+2^{m-1}\cdot\nu_{m-1}+\cdots+2^0\cdot\nu_{0}}
&\equiv\bs 01y.\nu_0(\cdots(\nu_{m-1}(1\, y)\cdots))
&(m\geq 1)
\end{align*}
that allow to encode the natural numbers in \textit{binary} notation.
Every word has type
$\BIntT\equiv\forall\alpha.!(\alpha\li\alpha)\li!(\alpha\li\alpha)\li\$(\alpha\li\alpha)$, where $m\geq 0$ and $\nu_{0\leq i\leq m-1}\in\{0,1\}$. Notice that every word is a Church numeral built using the two successors, identified by the variable names $0$, and $1$.
The combinators $\BSuccZ, \BSuccO$, and $\Pred$, with type $\BIntT\li\BIntT$, and
$\Branch$, with type $\BIntT\li\BIntT\li\BIntT\li\BIntT$, exist. They are the two successors, the predecessor and the branching, respectively. The branching yields its second argument as result, if the first one is the word $\BNum{0}$. Otherwise, the result is the third argument.
\paragraph{Eager tensor.}
We need the eager tensor to represent tuples of $\lambda$-terms. For every $m\geq 1$, the type \textit{eager tensor} is
$\bigodot^{m}_{i=1} A_i
\equiv
\forall \alpha.(\liv^{m}_{i=1} A_i\liv\alpha)\li\alpha$. Its type constructors coincide to the standard tuples in the $\lambda$-calculus:
\begin{align*}
\elan M_1\ldots M_m\eran
        &\equiv \bs z.z\,M_1\,\ldots\,M_m
        &(m\geq 1)
\\
\bs\elan x_1\ldots x_m\eran.M
        &\equiv \bs w.w(\bs x_1\ldots x_m. M)
        &(m\geq 1)
\end{align*}
Here, we can fairly assume that only to closed terms can be used in $\elan M_1\ldots M_m\eran$. In \cite{Roversi:2007-WALT-FULL} the constraint is a little bit more weak. So, for every closed $M_{1},\ldots,M_{m}$,
we have
$(\bs \elan x_1\cdots x_m\eran.M)\
\elan M_1,\ldots, M_m\eran
\rew^+
(\bs x_1 \ldots x_m.M)\,M_1\ldots M_m$.
\paragraph{Embedding.}
We can embed the arguments and the result of terms, with a functional type, into a suitable number of boxes. Since we have two kinds of implications, and we can transform the standard linear implication into an eager one, we have (at least) the following three kinds of embedding functors.
\par
The \textit{basic embedding} is $\BEmbed{n}{M}\equiv\bs x. M x$, for every $n\geq 1$.
It takes a term $M$, with type $L\li \S^{m} A$, for any $m\geq 0$, and yields one of type $\S^n L\liv \S^{m+n} A$.
\par
The \textit{linear embedding} is
$\LEmbed{n}{p}{M}\equiv\bs x_1 \ldots x_{p}. M\,x_1 \ldots x_{p}$,
for every $n,p\geq 0$. It takes a term $M$, with type $(\li^{p}_{i=1} L_i)\li\S^{m} A$,
for every $m\geq 0$, and yields one of type $(\li^{p}_{i=1}\S^n L_i)\li\S^{m+n} A$.
\par
The \textit{eager embedding}
$\EEmbed{n}{p}{q}{M}$ is:
$$\bs w_1\ldots w_p z_1\ldots z_q.
(\bs w_1\ldots w_p.M w_1\ldots w_p z_1\ldots z_q)
(\BEmbed{1}{\Coerc^n} w_1)
\ldots
(\BEmbed{1}{\Coerc^n} w_p),$$
for every $n, p, q\geq 0$.
It takes a term $M$ of type
$(\liv^{p}_{i=1} \S\BIntT)\liv(\liv^{q}_{j=1}\S^{m} L_j)\liv\S^{m} A$
and yields one of type
$(\liv^{p}_{i=1}\S^n \BIntT)\liv(\liv^{q}_{j=1}\S^{m+n} L_j)\liv\S^{m+n} A$.
\paragraph{Coercing.}
The \textit{coerce} function takes an instance of a binary word and reconstructs it inside a box. It is
$\Coerc\equiv\bs n. (\bs z. z\,\BNum{0})(n\, \BSuccZ\, \BSuccO)$.
To our purposes, $\Coerc$ must be iteratively composed to reconstruct a word into some given number of boxes:
\begin{align*}
\Coerc^{0}   &\equiv\bs x. x\\
\Coerc^{1}   &\equiv\Coerc\\
\Coerc^{m+1} &\equiv\bs x.\LEmbed{1}{1}{\Coerc^{m}}(\Coerc^1\,x)
&(m\geq 1)
\end{align*}
For every $m\geq 0$,
$\Coerc^{m}$ has type $\BIntT\li\$^{m}\BIntT$.
\paragraph{Eager diagonal.}
The \textit{eager diagonal} $\nabla^{m}_{n}$ is:
\begin{align*}
\bs w.
(\bs z.
 z\,\overbrace{\elan
               \BNum{0},\ldots,\BNum{0}
               \eran}^{n}
)
(w\, \!\!\begin{array}[t]{l}
     (\bs \elan x_1\ldots x_{n}\eran.
      \elan
      \BEmbed{m}{\BSuccZ}\, x_1,
      \ldots,
      \BEmbed{m}{\BSuccZ}\, x_{n}
      \eran
     )\\
     (\bs \elan x_1\ldots x_{n}\eran.
      \elan
      \BEmbed{m}{\BSuccO}\, x_1,
      \ldots,
      \BEmbed{m}{\BSuccO}\, x_{n}
      \eran
     )
 )
     \end{array}
\end{align*}
for every $m, n\geq 1$. It combines the copies of the word, given as its input, by means of an elementary tensor constructor. Namely,
$\nabla^{m}_{n}\, \BNum{a} \rew^+\overbrace{\elan\BNum{a},\ldots,\BNum{a}\eran}^{n}$.
Every copy is generated from scratch, by iterating the successors on words. The result is contained into a single box, but every component of the elementary tensor, in the result, is $m$ boxes deep. Namely,
$\nabla^{m}_{n}$ has type $\BIntT\li\S(\bigodot_{i=1}^{n}\S^{m}\BIntT)$.
\paragraph{Iterator.}
For every $\Intg{n}, \Intg{s} \geq 0$, and $m\geq 1$, and for every
closed term $G_0, G_1$, and $G_2$, all with type
$\$\BIntT\liv
     (\liv^{\Intg{n}}_{i=1}\$\BIntT)\liv
     (\liv^{\Intg{s}}_{j=1}\$^{m}\BIntT)\liv
     \$^{m}\BIntT\liv
     \$^{m}\BIntT$,
the iterator $\Iter{1+\Intg{n}}{\Intg{s}}{G_0}{G_1}{G_2}$ has type:
$$    \$\BIntT\liv
     (\liv^{\Intg{n}}_{i=1}\$\BIntT)\liv
     (\liv^{\Intg{s}}_{i=1}\$^{m+4}\BIntT)
     \liv
     \$^{m+4}\BIntT$$
As expected, the first argument of $\Iter{1+\Intg{n}}{\Intg{s}}{G_0}{G_1}{G_2}$ drives the iteration, $G_2$ is the base function, while $G_0$ and $G_1$ are the inductive ones, chosen by the ``bits'' in the first argument itself of
$\Iter{1+\Intg{n}}{\Intg{s}}{G_0}{G_1}{G_2}$.
\par
To formally recall the behavior of the iterator, we need to assume that:
(i)
$\Intg{n}, \Intg{s}\geq 0$,
$\BNum{a},
 \BNum{n},
 \BNum{n_{1}},\ldots,\BNum{n_{\Intg{n}}},
 \BNum{s_{1}},\ldots,\BNum{s_{\Intg{s}}}$ be some words,
(ii) $\{\nu_0, \nu_1, \ldots\}$ be a denumerable set of metavariables to range over $\{0, 1\}$,
(iii)
$[\BNum{x}]^i$ denotes a list with $i\in\Nat$ copies of the word $\BNum{x}$, for any $x$.
\par
Also, we assume that:
\begin{itemize}
   \item
    $G_{2}\,
    \BNum{0}\,
    \BNum{n_{1}}\,\ldots\,\BNum{n_{\Intg{n}}}\,
    \BNum{s_{1}}\,\ldots\,\BNum{s_{\Intg{s}}}\,
    \BNum{0}$ rewrites to a word $\BNum{a}$, and
   \item
    $G_{1}\,
    \BNum{0}\,
    \BNum{n_{1}}\,\ldots\,\BNum{n_{\Intg{n}}}\,
    \BNum{s_{1}}\,\ldots\,\BNum{s_{\Intg{s}}}\,
    \BNum{a}$ rewrites to a word, denoted as
    $r[0,a,n_1,\ldots,n_{\Intg{n}},s_1,\ldots,s_{\Intg{s}}]$, and
   \item
    for every $m, i$, such that $m\geq0, m-1\geq i\geq 0$,
    \small
    $$G_{\nu_i}\,
    \BNum{\left(\sum_{j=0}^{m-(i+1)} 2^{m-(i+1)-j} \nu_{m-j}\right)}\,
    \BNum{n_{1}}\,\ldots\,\BNum{n_{\Intg{n}}}\,
    \BNum{s_{1}}\,\ldots\,\BNum{s_{\Intg{s}}}\,
    r[m-(i+1),a,n_1,\ldots,n_{\Intg{n}},s_1,\ldots,s_{\Intg{s}}]
    $$
    \normalsize
    rewrites to a word, denoted as
    $r[m-i,a,n_1,\ldots,n_{\Intg{n}},s_1,\ldots,s_{\Intg{s}}]$.
\end{itemize}
Then:
\small
\begin{align}
\nonumber
\Iter{1+\Intg{n}}{\Intg{s}}{G_0}{G_1}{G_2}\,
\BNum{0}\,
\BNum{n_{1}}\,\ldots\,\BNum{n_{\Intg{n}}}\,
\BNum{s_{1}}\,\ldots\,\BNum{s_{\Intg{s}}}
&\rew^+ \BNum{a}
\\
\nonumber
\Iter{1+\Intg{n}}{\Intg{s}}{G_0}{G_1}{G_2}\,
\BNum{\left(\sum_{j=0}^{m} 2^{j} \nu_j \right)}\,
\BNum{n_{1}}\,\ldots\,\BNum{n_{\Intg{n}}}\,
\BNum{s_{1}}\,\ldots\,\BNum{s_{\Intg{s}}}
&\rew^+
r[m,a,n_1,\ldots,n_{\Intg{n}},s_1,\ldots,s_{\Intg{s}}]
&\left(\BNum{\sum_{j=0}^{m} 2^{j} \nu_j}\neq 0\right)
\end{align}
\normalsize
The full details about $\Iter{1+\Intg{n}}{\Intg{s}}{G_0}{G_1}{G_2}$ are in \cite{Roversi:2007-WALT-FULL}, whose keypoint is to prove that such a combinator is indeed representable inside \WALT, giving its completeness w.r.t. \QlSRN.

%% file: figure-of-the-type-assignment.tex
\begin{figure}
\begin{center}
\fbox{
\begin{minipage}{.9\textwidth}
\begin{tabular}{c}
 \infer[A]
  {
   \Gamma,\ta{x}{L};\Delta;{\mathcal E}\vdash \ta{x}{L}
  }
  {
  }
\\
\\
 \infer[C]
  {
   \Gamma;
   \Delta;
   {\mathcal E}\sqcup
   \{(\Theta_x,\Theta_y;\{\ta{z}{A}\})\}
   \vdash \ta{M\{^{z}/_{x} {}^{z}/_{y}\}}{B}
  }
  {
   \Gamma; \Delta;
   {\mathcal E},
   (\Theta_x;\{\ta{x}{A}\}),
   (\Theta_y;\{\ta{y}{A}\})
   \vdash \ta{M}{B}
  }
\\
\\
 \infer[\li I]
  {\Gamma; \Delta; {\mathcal E}
   \vdash \ta{\bs x.M}{L\li B}}
  {\Gamma,\ta{x}{L}; \Delta; {\mathcal E}\vdash \ta{M}{B}}
\qquad
 \infer[\li I_{\$}]
  {\Gamma; \Delta; {\mathcal E}
   \vdash \ta{\bs x.M}
             {\$A\li B}}
  {\Gamma;\Delta,\ta{x}{A}; {\mathcal E}\vdash \ta{M}{B}}
\\
\\
 \infer[\li E]
  {\Gamma_M, \Gamma_N;
   \Delta_M, \Delta_N;
   {\mathcal E}_M
   \sqcup
   {\mathcal E}_N
   \vdash \ta{MN}{B}
  }
  {\Gamma_M;
   \Delta_M;
   {\mathcal E}_M
   \vdash \ta{M}{A\li B}
   &
   \Gamma_N;
   \Delta_N;
   {\mathcal E}_N
   \vdash \ta{N}{A}
   &
   A\not\equiv !C, \text{ for any } C
   }
\\
\\
 \infer[\li I_!]
  {
   \Gamma;
   \Delta;
   {\mathcal E}\sqcup
   \{(\Theta;\emptyset)\}
   \vdash \ta{\bs x.M}{\ !A\li B}
  }
  {
   \Gamma; \Delta;
   {\mathcal E},
   (\Theta;\{\ta{x}{A}\})
   \vdash \ta{M}{B}
  }
\\
\\
 \infer[\li E_!]
  {\Gamma_M, \Gamma_N;
   \Delta_M, \Delta_N;
   {\mathcal E}_M
   \sqcup
   {\mathcal E}_N
   \vdash \ta{MN}{B}
  }
  {
   \Gamma_M;
   \Delta_M;
   {\mathcal E}_M
   \vdash \ta{M}{\,!A\li B}
   &
   \Gamma_N;
   \Delta_N;
   {\mathcal E}_N
   \vdash \ta{N}{\,!A}
   &
   {\mathcal E}_M\subseteq\{(\emptyset;\Phi_1),\ldots,(\emptyset;\Phi_n)\}
}
\\
\\
 \infer[\liv I]
  {
   \Gamma;\Delta;{\mathcal E}\sqcup\{(\Theta;\emptyset)\}
   \vdash \ta{\bs x.M}{\$A\liv B}
  }
  {
   \Gamma;\Delta;
   {\mathcal E},(\Theta,\ta{x}{A}; \emptyset)
   \vdash \ta{M}{B}
  }
\\
\\
\infer[\liv E]
  {
   \Gamma_M;
   \Delta;
   {\mathcal E}_M
   \sqcup
   {\mathcal E}_N
   \vdash \ta{MN}{B}
  }
  {
   \Gamma_M;
   \Delta;
   {\mathcal E}_M
   \vdash \ta{M}{\$A\liv B}
   &
   \emptyset;
   \emptyset;
   {\mathcal E}_N
   \vdash \ta{N}{\$A}
   &
   {\mathcal E}_N\subseteq\{(\Theta;\emptyset)\}
   }
\\
\\
\infer[\$]
{
 \Gamma';
 \$\Delta',\Delta;
 \{(\$\Theta';\emptyset)\}
 \sqcup
 \{(\Theta_1;\Phi_1)\}
 \sqcup\ldots\sqcup
 \{(\Theta_m;\Phi_m)\}
 \vdash \ta{M}{\$B}
}
{
 \Gamma;
 \Delta';
 \{(\Theta';\emptyset)\}
 \vdash \ta{M}{B}
 &
 \Gamma\subseteq
 \Delta\cup\bigcup_{i=1}^{m}\Theta_i
 \cup\bigcup_{i=1}^{m}\Phi_i
 &
 \Theta_i\neq\emptyset\text{ iff }\Phi_i=\emptyset
}
\\
\\
\infer[!]
 {\Gamma';
  \Delta;
  \{(\$\Theta';\emptyset)\}\sqcup\{(\Theta;\Phi)\}
  \vdash \ta{M}{\ !B}
 }
 {
  \Gamma;
  \emptyset;
  \{(\Theta';\emptyset)\}
  \vdash \ta{M}{B}
  &
  \Gamma
  \subseteq
  \Theta\cup\Phi
  &
  \Theta\neq\emptyset \Rightarrow \dom{\Phi}\cap\FV{M}\neq\emptyset
 }
\\
\\
 \infer[\forall I]
  {\Gamma; \Delta; {\mathcal E}\vdash \ta{M}{\forall \alpha.L}}
  {
  \Gamma; \Delta; {\mathcal E}\vdash \ta{M}{L}
  &
  \text{$\alpha$ not free in $\Gamma, \Delta$ and ${\mathcal E}$}
  }

\qquad
 \infer[\forall E]
  {\Gamma; \Delta; {\mathcal E}\vdash \ta{M}{L\{L'/\alpha\}}}
  {\Gamma; \Delta; {\mathcal E}\vdash \ta{M}{\forall\alpha.L}}
\end{tabular}
\end{minipage}
}\end{center}
\normalsize
\caption{Weak Affine Light Typing}
\label{figure:Weak LAL as a type assignment}
\end{figure}

%% file: canonical-bang-box.pstex_t
\begin{picture}(0,0)%
\includegraphics{canonical-bang-box.pstex}%
\end{picture}%
\setlength{\unitlength}{4144sp}%
\begingroup\makeatletter\ifx\SetFigFontNFSS\undefined%
\gdef\SetFigFontNFSS#1#2#3#4#5{%
  \reset@font\fontsize{#1}{#2pt}%
  \fontfamily{#3}\fontseries{#4}\fontshape{#5}%
  \selectfont}%
\fi\endgroup%
\begin{picture}(1650,1881)(616,-940)
\put(1036,-691){\makebox(0,0)[lb]{\smash{{\SetFigFontNFSS{12}{14.4}{\familydefault}{\mddefault}{\updefault}{\color[rgb]{0,0,0}$\ldots$}%
}}}}
\put(1936,-871){\makebox(0,0)[lb]{\smash{{\SetFigFontNFSS{12}{14.4}{\familydefault}{\mddefault}{\updefault}{\color[rgb]{0,0,0}$\tc{blue}{!B}$}%
}}}}
\put(1441,704){\makebox(0,0)[lb]{\smash{{\SetFigFontNFSS{12}{14.4}{\familydefault}{\mddefault}{\updefault}{\color[rgb]{0,0,0}$!A$}%
}}}}
\put(1441,389){\makebox(0,0)[lb]{\smash{{\SetFigFontNFSS{12}{14.4}{\familydefault}{\mddefault}{\updefault}{\color[rgb]{0,0,0}$A$}%
}}}}
\put(1396,-331){\makebox(0,0)[lb]{\smash{{\SetFigFontNFSS{12}{14.4}{\familydefault}{\mddefault}{\updefault}{\color[rgb]{0,0,0}$C_n$}%
}}}}
\put(1756,-331){\makebox(0,0)[lb]{\smash{{\SetFigFontNFSS{12}{14.4}{\familydefault}{\mddefault}{\updefault}{\color[rgb]{0,0,0}$B$}%
}}}}
\put(2251,434){\makebox(0,0)[lb]{\smash{{\SetFigFontNFSS{12}{14.4}{\familydefault}{\mddefault}{\updefault}{\color[rgb]{0,0,0}$!$}%
}}}}
\put(856,-331){\makebox(0,0)[lb]{\smash{{\SetFigFontNFSS{12}{14.4}{\familydefault}{\mddefault}{\updefault}{\color[rgb]{0,0,0}$C_1$}%
}}}}
\put(1396,-871){\makebox(0,0)[lb]{\smash{{\SetFigFontNFSS{12}{14.4}{\familydefault}{\mddefault}{\updefault}{\color[rgb]{0,0,0}$\tc{red}{\S C_n}$}%
}}}}
\put(631,-871){\makebox(0,0)[lb]{\smash{{\SetFigFontNFSS{12}{14.4}{\familydefault}{\mddefault}{\updefault}{\color[rgb]{0,0,0}$\tc{red}{\S C_1}$}%
}}}}
\put(1306, 29){\makebox(0,0)[lb]{\smash{{\SetFigFontNFSS{12}{14.4}{\familydefault}{\mddefault}{\updefault}{\color[rgb]{0,0,0}$\Pi_M$}%
}}}}
\end{picture}%

%% file: eager-arrow-elim.pstex_t
\begin{picture}(0,0)%
\includegraphics{eager-arrow-elim.pstex}%
\end{picture}%
\setlength{\unitlength}{4144sp}%
\begingroup\makeatletter\ifx\SetFigFontNFSS\undefined%
\gdef\SetFigFontNFSS#1#2#3#4#5{%
  \reset@font\fontsize{#1}{#2pt}%
  \fontfamily{#3}\fontseries{#4}\fontshape{#5}%
  \selectfont}%
\fi\endgroup%
\begin{picture}(1872,3759)(169,-4618)
\put(1036,-2941){\makebox(0,0)[lb]{\smash{{\SetFigFontNFSS{12}{14.4}{\familydefault}{\mddefault}{\updefault}{\color[rgb]{0,0,0}\scriptsize$\liv\!\!I$}%
}}}}
\put(2026,-2356){\makebox(0,0)[lb]{\smash{{\SetFigFontNFSS{12}{14.4}{\familydefault}{\mddefault}{\updefault}{\color[rgb]{0,0,0}$\S$}%
}}}}
\put(1666,-2086){\makebox(0,0)[lb]{\smash{{\SetFigFontNFSS{12}{14.4}{\familydefault}{\mddefault}{\updefault}{\color[rgb]{0,0,0}$\tc{red}{\S C'}$}%
}}}}
\put(1666,-2716){\makebox(0,0)[lb]{\smash{{\SetFigFontNFSS{12}{14.4}{\familydefault}{\mddefault}{\updefault}{\color[rgb]{0,0,0}$\tc{red}{\S \Theta'}$}%
}}}}
\put(1036,-1951){\makebox(0,0)[lb]{\smash{{\SetFigFontNFSS{12}{14.4}{\familydefault}{\mddefault}{\updefault}{\color[rgb]{0,0,0}\scriptsize$\liv\!\!E$}%
}}}}
\put(1036,-1321){\makebox(0,0)[lb]{\smash{{\SetFigFontNFSS{12}{14.4}{\familydefault}{\mddefault}{\updefault}{\color[rgb]{0,0,0}\scriptsize$\liv\!\!I$}%
}}}}
\put(1666,-3706){\makebox(0,0)[lb]{\smash{{\SetFigFontNFSS{12}{14.4}{\familydefault}{\mddefault}{\updefault}{\color[rgb]{0,0,0}$\tc{red}{\S C}$}%
}}}}
\put(2026,-3976){\makebox(0,0)[lb]{\smash{{\SetFigFontNFSS{12}{14.4}{\familydefault}{\mddefault}{\updefault}{\color[rgb]{0,0,0}$\S$}%
}}}}
\put(1036,-3571){\makebox(0,0)[lb]{\smash{{\SetFigFontNFSS{12}{14.4}{\familydefault}{\mddefault}{\updefault}{\color[rgb]{0,0,0}\scriptsize$\liv\!\!E$}%
}}}}
\put(1891,-4336){\makebox(0,0)[lb]{\smash{{\SetFigFontNFSS{12}{14.4}{\familydefault}{\mddefault}{\updefault}{\color[rgb]{0,0,0}$\tc{red}{\S \Theta}$}%
}}}}
\put(1486,-4336){\makebox(0,0)[lb]{\smash{{\SetFigFontNFSS{12}{14.4}{\familydefault}{\mddefault}{\updefault}{\color[rgb]{0,0,0}$\tc{red}{\$C'}$}%
}}}}
\put(496,-3931){\makebox(0,0)[lb]{\smash{{\SetFigFontNFSS{12}{14.4}{\familydefault}{\mddefault}{\updefault}{\color[rgb]{0,0,0}$\tc{red}{\S C}\!\liv\! A$}%
}}}}
\put(451,-2311){\makebox(0,0)[lb]{\smash{{\SetFigFontNFSS{12}{14.4}{\familydefault}{\mddefault}{\updefault}{\color[rgb]{0,0,0}$\tc{red}{\S C'}\!\liv\! A$}%
}}}}
\put(1576,-2401){\makebox(0,0)[lb]{\smash{{\SetFigFontNFSS{12}{14.4}{\familydefault}{\mddefault}{\updefault}{\color[rgb]{0,0,0}$\Pi_{N}$}%
}}}}
\put(1576,-4021){\makebox(0,0)[lb]{\smash{{\SetFigFontNFSS{12}{14.4}{\familydefault}{\mddefault}{\updefault}{\color[rgb]{0,0,0}$\Pi_{M}$}%
}}}}
\end{picture}%

%% file: eager-and-modal-arrow.pstex_t
\begin{picture}(0,0)%
\includegraphics{eager-and-modal-arrow.pstex}%
\end{picture}%
\setlength{\unitlength}{4144sp}%
\begingroup\makeatletter\ifx\SetFigFontNFSS\undefined%
\gdef\SetFigFontNFSS#1#2#3#4#5{%
  \reset@font\fontsize{#1}{#2pt}%
  \fontfamily{#3}\fontseries{#4}\fontshape{#5}%
  \selectfont}%
\fi\endgroup%
\begin{picture}(2010,3411)(211,-2740)
\put(1396,-1051){\makebox(0,0)[lb]{\smash{{\SetFigFontNFSS{12}{14.4}{\familydefault}{\mddefault}{\updefault}{\color[rgb]{0,0,0}$!A$}%
}}}}
\put(1396,-1366){\makebox(0,0)[lb]{\smash{{\SetFigFontNFSS{12}{14.4}{\familydefault}{\mddefault}{\updefault}{\color[rgb]{0,0,0}$A$}%
}}}}
\put(1711,-2086){\makebox(0,0)[lb]{\smash{{\SetFigFontNFSS{12}{14.4}{\familydefault}{\mddefault}{\updefault}{\color[rgb]{0,0,0}$B$}%
}}}}
\put(2206,-1321){\makebox(0,0)[lb]{\smash{{\SetFigFontNFSS{12}{14.4}{\familydefault}{\mddefault}{\updefault}{\color[rgb]{0,0,0}$!$}%
}}}}
\put(226,-2671){\makebox(0,0)[lb]{\smash{{\SetFigFontNFSS{12}{14.4}{\familydefault}{\mddefault}{\updefault}{\color[rgb]{0,0,0}$\tc{red}{\S C}$}%
}}}}
\put(586,-556){\makebox(0,0)[lb]{\smash{{\SetFigFontNFSS{12}{14.4}{\familydefault}{\mddefault}{\updefault}{\color[rgb]{0,0,0}\scriptsize$\li\!I_!$\normalsize}%
}}}}
\put(586,254){\makebox(0,0)[lb]{\smash{{\SetFigFontNFSS{12}{14.4}{\familydefault}{\mddefault}{\updefault}{\color[rgb]{0,0,0}\scriptsize$\liv\!\!I$\normalsize}%
}}}}
\put(1711,-2491){\makebox(0,0)[lb]{\smash{{\SetFigFontNFSS{12}{14.4}{\familydefault}{\mddefault}{\updefault}{\color[rgb]{0,0,0}$\tc{blue}{!B}$}%
}}}}
\put(946,-2086){\makebox(0,0)[lb]{\smash{{\SetFigFontNFSS{12}{14.4}{\familydefault}{\mddefault}{\updefault}{\color[rgb]{0,0,0}$C$}%
}}}}
\end{picture}%

%% file: modal-arrow-elimination.pstex_t
\begin{picture}(0,0)%
\includegraphics{modal-arrow-elimination.pstex}%
\end{picture}%
\setlength{\unitlength}{4144sp}%
\begingroup\makeatletter\ifx\SetFigFontNFSS\undefined%
\gdef\SetFigFontNFSS#1#2#3#4#5{%
  \reset@font\fontsize{#1}{#2pt}%
  \fontfamily{#3}\fontseries{#4}\fontshape{#5}%
  \selectfont}%
\fi\endgroup%
\begin{picture}(2536,3276)(2371,-7420)
\put(3781,-5776){\makebox(0,0)[lb]{\smash{{\SetFigFontNFSS{12}{14.4}{\familydefault}{\mddefault}{\updefault}{\color[rgb]{0,0,0}$!A$}%
}}}}
\put(3781,-6091){\makebox(0,0)[lb]{\smash{{\SetFigFontNFSS{12}{14.4}{\familydefault}{\mddefault}{\updefault}{\color[rgb]{0,0,0}$A$}%
}}}}
\put(4096,-6811){\makebox(0,0)[lb]{\smash{{\SetFigFontNFSS{12}{14.4}{\familydefault}{\mddefault}{\updefault}{\color[rgb]{0,0,0}$B$}%
}}}}
\put(4591,-6046){\makebox(0,0)[lb]{\smash{{\SetFigFontNFSS{12}{14.4}{\familydefault}{\mddefault}{\updefault}{\color[rgb]{0,0,0}$!$}%
}}}}
\put(2881,-7351){\makebox(0,0)[lb]{\smash{{\SetFigFontNFSS{12}{14.4}{\familydefault}{\mddefault}{\updefault}{\color[rgb]{0,0,0}$\tc{red}{\S D}$}%
}}}}
\put(3646,-4336){\makebox(0,0)[lb]{\smash{{\SetFigFontNFSS{12}{14.4}{\familydefault}{\mddefault}{\updefault}{\color[rgb]{0,0,0}$C$}%
}}}}
\put(3376,-6811){\makebox(0,0)[lb]{\smash{{\SetFigFontNFSS{12}{14.4}{\familydefault}{\mddefault}{\updefault}{\color[rgb]{0,0,0}$D$}%
}}}}
\put(3466,-4606){\makebox(0,0)[lb]{\smash{{\SetFigFontNFSS{12}{14.4}{\familydefault}{\mddefault}{\updefault}{\color[rgb]{0,0,0}\scriptsize$\li\!\!E_!$}%
}}}}
\put(4096,-7081){\makebox(0,0)[lb]{\smash{{\SetFigFontNFSS{12}{14.4}{\familydefault}{\mddefault}{\updefault}{\color[rgb]{0,0,0}$\tc{blue}{!B}$}%
}}}}
\put(2386,-5191){\makebox(0,0)[lb]{\smash{{\SetFigFontNFSS{12}{14.4}{\familydefault}{\mddefault}{\updefault}{\color[rgb]{0,0,0}$\Pi_{M}$}%
}}}}
\put(2971,-4921){\makebox(0,0)[lb]{\smash{{\SetFigFontNFSS{12}{14.4}{\familydefault}{\mddefault}{\updefault}{\color[rgb]{0,0,0}$\tc{blue}{!B}\liv C$}%
}}}}
\put(3466,-5281){\makebox(0,0)[lb]{\smash{{\SetFigFontNFSS{12}{14.4}{\familydefault}{\mddefault}{\updefault}{\color[rgb]{0,0,0}\scriptsize$\li\!I_!$\normalsize}%
}}}}
\put(4321,-5281){\makebox(0,0)[lb]{\smash{{\SetFigFontNFSS{12}{14.4}{\familydefault}{\mddefault}{\updefault}{\color[rgb]{0,0,0}$\Pi_{N}$}%
}}}}
\put(4456,-4921){\makebox(0,0)[lb]{\smash{{\SetFigFontNFSS{12}{14.4}{\familydefault}{\mddefault}{\updefault}{\color[rgb]{0,0,0}$\tc{blue}{! B}$}%
}}}}
\put(3421,-5551){\makebox(0,0)[lb]{\smash{{\SetFigFontNFSS{12}{14.4}{\familydefault}{\mddefault}{\updefault}{\color[rgb]{0,0,0}$C$}%
}}}}
\end{picture}%

%% file: safe-recursion-on-notation.tex
\section{Safe Recursion on Notation}
\label{section:Safe Recursion on Notation}
We recall two classes of functions: \textit{Safe recursion on notation} (\SRN) \cite{Bellantoni92CC}, and \textit{Composition-linear safe recursion on notation} (\QlSRN), both in the style of \cite{Beckmann:96-AML}.
Remark that Composi\-tion-linear safe recursion on notation was identified as \textit{Quasi-linear safe recursion on notation} in \cite{Roversi:2007-WALT-FULL}. The reason for the name changing will be given in the conclusions (Section~\ref{section:Conclusions and future work}).
\paragraph{The signature of Safe recursion on notation.}
Let $\Sigma_{\SRN}=\cup_{k,l\in\Nat} \Sigma^{k,l}_{\SRN}$ be the
signature of Safe recursion on notation .
$\Sigma_{\SRN}$ contains the \textit{base functions} and it is closed under the schemes called \textit{safe composition} and \textit{safe recursion}.
For every $k,l\in\Nat$, the \textit{base functions} are
the \textit{zero} $\zero{k}{l}\in\Sigma^{k;l}_{\SRN}$,
the \textit{successors} $\sucz, \suco$,
the \textit{predecessor} $\pred\in\Sigma^{0;1}_{\SRN}$,
the \textit{projection} $\proj{k}{l}{i}\in\Sigma^{k;l}_{\SRN}$, with $1\leq i\leq k+l$, and the \textit{branching} $\bran\in\Sigma^{0;3}_{\SRN}$.
\par
The \textit{safe composition} is
$\comp{k}
      {l}
      {k'}
      {l'}
      {f,g_1,\ldots,g_{k'}
	,h_1,\ldots,h_{l'}
      }\in\Sigma^{k;l}_{\SRN}$
if
$f\in\Sigma^{k';l'}_{\SRN}$,
$g_1,\ldots,g_{k'}\in\Sigma^{k;0}_{\SRN}$, and
$h_{i}\in\Sigma^{k;l}_{\SRN}$, for every $k,k',l,l'\in\Nat$.
\par
The \textit{safe recursion} is
$\rec{k+1}{l}{g,h_0,h_1}\in\Sigma^{k+1;l}_{\SRN}$
if
$g\in\SRN^{k;l}$, and
$h_0,h_1\in\Sigma^{k+1;l+1}_{\SRN}$.
\paragraph{Safe recursion on notation.}
Let $\SRNVnames$ be a denumerable set of \emph{names of variables}, disjoint from $\Sigma_{\SRN}$. The set Safe recursion on notation (\SRN) contains functions with signature $\Sigma_{\SRN}$. \SRN\ is defined as follows. $\SRNVnames\subset\SRN$, and
for every $k,l\in\Nat$,
if
$f\in\Sigma_{\SRN}^{k;l}$, and
$t_1,\ldots,t_{k},u_1,\ldots,u_{l}\in\SRN$,
then $f(t_1,\ldots,t_{k},u_1,\ldots,u_{l})\in\SRN$. A term is \textit{closed} if it does not contain variables of $\SRNVnames$.
\paragraph{Notations and terminology.}
$x, y, z \ldots$ denote elements of $\SRNVnames$.
$t, u, v \ldots$ denote elements of $\SRN$.
For every $f\in\Sigma^{k,l}_{\SRN}$, $k$ and $l$ are the \textit{normal} and \textit{safe} arity of $f$, respectively.
For every $k, l\in\Nat$, such that $l-k\geq1$,
$\seq{t}{k}{l}$ denotes a \textit{non empty} sequence
$t_k,\ldots,t_l$ of $l-k+1$ terms in \SRN.
$\seq{t}{k}{l}(i)$, with $k\leq i\leq l$, denotes the element $t_i$
of $\seq{t}{k}{l}$.
\paragraph{An equational theory on \SRN.}
The definition of the equational theory exploits that every natural number $n$ can be written, uniquely, in binary notation, as $\sum^{m}_{j=0}2^{m-j} \nu_{m-j}$.
So, assuming to abbreviate the base functions $\sucz, \suco$ as
$\Sucz, \Suco$, respectively, we can follow \cite{MurawskiOng00} and say that $0$ is equivalent to $\zero{0}{0}$, and $n\geq 1$ to
$\mathtt{s}_{\nu_0}(\ldots(\mathtt{s}_{\nu_{m-1}}(\Suco\,\zero{0}{0}))\ldots)$.
Then, the \textit{equational theory} is as follows.
\textit{Zero} is constantly equal to $0$:
$\zero{k}{l}(\seq{x}{1}{k},\seq{x}{k+1}{k+l})=\Zero$
for any $k,l\in\Nat$.
The \textit{predecessor} erases the least significant bit of any number greater than $0$:
for every $i\in\{0,1\}$, $\pred(\Zero)= \Zero$, and $\pred(\Suc{i}(y))=y$.
We shall use $\preds$ as an abbreviation of $\pred$.
The \textit{conditional} has three arguments. If the first is zero, then the result is the second argument. Otherwise, it is the third one:
for every $i\in\{0,1\}$, $\bran(\Zero,y_0,y_1) = y_0$, and $\bran(\Suc{i}(y),y_0,y_1) = y_1$.
The \textit{projection} chooses one argument, out of a given tuple, as a result: for every $1\leq i\leq k+l$, $\proj{k}{l}{i}(\seq{x}{1}{k},\seq{x}{k+1}{k+l}) = x_i$.
\par
The \textit{safe composition} uses as arguments of $f$ both the results of the \textit{normal functions} $g_1,\ldots,g_{k'}$, applied to $k$ normal arguments, and the result of the \textit{safe functions} $h_1,\ldots,h_{l'}$, applied to $k$ normal and $l$ safe arguments:
\small
\begin{alignat*}{1}
&
\comp{k}
     {l}
     {k'}
     {l'}
     {f,g_1,\ldots,g_{k'},h_1,\ldots,h_{l'}}
(\seq{x}{1}{k}
,\seq{x}{k+1}{k+l})
\\
&
\qquad
\qquad
= f\!\!
  \begin{array}[t]{l}
  (g_1(\seq{x}{1}{k})
  ,\ldots
  ,g_{k'}(\seq{x}{1}{k})
  ,h_1(\seq{x}{1}{k},\seq{x}{k+1}{k+l})
  ,\ldots
  ,h_{l'}(\seq{x}{1}{k},\seq{x}
                            {k+1}
                            {k+l}))
\enspace .
  \end{array}
\end{alignat*}
\normalsize
The \textit{recursion} iterates either the function $h_0$, or $h_1$, as many times as
the length of its first argument. The choice between $h_0$, and $h_1$ depends on the
least significant digit of the first argument, while
the base of the iteration is a function $g$. The recursion is:
\small
\begin{alignat*}{1}
\rec{k+1}{l}{g,h_0,h_1}
 (\Zero
 ,\seq{x}{1}{k}
 ,\seq{x}{k+1}{k+l}
 )
& =
 g(\seq{x}{1}{k},\seq{x}{k+1}{k+l})
 \\
\rec{k+1}{l}{g,h_0,h_1}
(\Suc{i}(x)
,\seq{x}{1}{k}
,\seq{x}{k+1}{k+l}
)
&=
h_i(x
   ,\seq{x}{1}{k}
   ,\seq{x}{k+1}{k+l}
   ,\rec{k+1}{l}{g,h_0,h_1}(x
                           ,\seq{x}{1}{k}
                           ,\seq{x}{k+1}{k+l}))
\enspace .
\end{alignat*}
\normalsize
\subsection{Composition-linear safe recursion on notation}
\QlSRN\ is \SRN\ with a restricted form of safe composition, which we call \textit{linear (safe) composition}.
Its signature is
$\lcomp{k}
       {\sum_{i=1}^{l'}l_i}
       {k'}
       {l'}
       {f,g_1,\ldots,g_{k'}
        ,h_1,\ldots,h_{l'}
       }\in\Sigma^{k;\sum_{i=1}^{l'}l_i}_{\QlSRN}$
if
$f\in\Sigma^{k';l'}_{\QlSRN}$,
$g_1,\ldots,g_{k'}\in\Sigma^{k;0}_{\QlSRN}$, and
$h_{i}\in\Sigma^{k;l_i}_{\QlSRN}$, with
$i\in\{1,\ldots,l'\}$, for every $k,l,l',l_1,\ldots,l_{l'}\in\Nat$.
Namely, unlike the general safe composition scheme of \SRN, the safe arguments are used linearly: the list of safe arguments is split into as many sub-sequences as required by the safe arity of every safe function $h_j$, with $1\leq j\leq l'$:
\small
\begin{alignat*}{1}
&
\lcomp{k}
      {\sum_{i=1}^{l'} l_i}
      {k'}
      {l'}
      {f,g_1,\ldots,g_{k'},h_1,\ldots,h_{l'}}
\!\!
\begin{array}[t]{l}
(\seq{x}{1}{k}
,\seq{x}{k+1}{k+l_1},\seq{x}{k+1+l_1}{k+l_1+l_2}
,\ldots,\seq{x}{k+1+\sum_{i=1}^{l'-1}l_i}
               {k+1+\sum_{i=1}^{l'}l_i})
\end{array}
\\
&
\qquad
\qquad
= f\!\!
  \begin{array}[t]{l}
  (g_1(\seq{x}{1}{k})
  ,\ldots
  ,g_{k'}(\seq{x}{1}{k})
  ,h_1(\seq{x}{1}{k},\seq{x}{k+1}{k+l_1})
  ,\ldots
  ,h_{l'}(\seq{x}{1}{k},\seq{x}
                            {k+1+\sum_{i=1}^{l'-1}l_i}
                            {k+1+\sum_{i=1}^{l'}l_i}))
\enspace .
  \end{array}
\end{alignat*}
\normalsize

%% file: square-composition.tex
\section{The full safe composition of \SRN\ in \WALT}
\label{section:The safe composition in WALT}
We know that \WALT\ contains \QlSRN\ as its subsystem. Namely, it holds:
\begin{theorem}[\cite{Roversi:2007-WALT-FULL}.]
There is a map $\srtw{\, }{}$, such that, if $f(n_1\ldots,n_{k},s_1\ldots,s_{l})$ belongs to \QlSRN, and $f(n_1,\ldots,n_k,s_1,$ $\ldots,s_l)=n$, then
$\srtw{f(n_1,\ldots,n_k,s_1,\ldots,s_l)}{}\rew^*\BNum{n}$,
for every $n_1,\ldots,n_k,s_1,\ldots, s_l$, $n\in\Nat$, in binary notation.
\end{theorem}
Here we go further by defining the combinators that, using the base combinators of Section~\ref{subsection:Basic combinators}, give the \textit{full} safe composition scheme of \SRN\ as a term of \WALT.
The definitions here below will realize the functional blocks of the example in Figure~\ref{figure:square-composition-block-diagram}.
\paragraph{Sharing safe names.}
Let $\Intg{n}, \Intg{s}\geq 0$, $m\geq 1$, and a closed term $M$, with type
$(\liv^{\Intg{n}}_{i=1}\S\BIntT)
  \liv
  (\liv^{\Intg{s}+1}_{j=1}\S^{m}\BIntT)
  \liv
 \S^{m}\BIntT$, be given.
We call $\share{\Intg{n}}{\Intg{s}}{m}$ the closed term that takes $M$ and $\Intg{n}+\Intg{s}$ words as its arguments. The first $\Intg{n}$ arguments can be viewed as a normal ones, while the $\Intg{s}$ second ones as safe.
Then, $\share{\Intg{n}}{\Intg{s}}{m}$ applies $M$ to the normal and safe arguments, using $\BNum{s_{\Intg{s}}}$ as a value for the two last safe positions.
Namely, the \textit{behavior} is:
$$\share{\Intg{n}}{\Intg{s}}{m}[M]
\BNum{n_1}\ldots\BNum{n_{\Intg{n}}}\,\BNum{s_1}\ldots\BNum{s_{\Intg{s}}}
\rew^+
M\,\BNum{n_1}\ldots\BNum{n_{\Intg{n}}}\,
\BNum{s_1}\ldots\BNum{s_{\Intg{s}}}\,\BNum{s_{\Intg{s}}}$$
Clearly, using the same safe value twice, after its duplication, we are sharing it.
\par
The \textit{type} of $\share{\Intg{n}}{\Intg{s}}{m}$ is:
\[
((\liv^{\Intg{n}}_{i=1}\S\BIntT)
  \liv
  (\liv^{\Intg{s}+1}_{j=1}\S^{m}\BIntT)
  \liv
 \S^{m}\BIntT)
 \liv
(\liv^{\Intg{n}}_{i=1}\S\BIntT)
 \liv
 (\liv^{\Intg{s}}_{j=1}\S^{m+4}\BIntT)
 \liv
\S^{m+4}\BIntT
\]
\par
The \textit{definition} of $\share{\Intg{n}}{\Intg{s}}{m}$ is:
$$
\Iter
{1+\Intg{n}}
{\Intg{s}}
{\bs w\,x_1\ldots x_{\Intg{n}}\,y_1\ldots y_{\Intg{s}+1}.\BNum{0}}
{\bs w.M}
{\bs w\,
     x_1\ldots x_{\Intg{n}}\,
     y_1\ldots y_{\Intg{s}+1}. y_{\Intg{s}}}
\,\BNum{1}$$
\paragraph{Rotating safe names.}
Let $\Intg{n}, \Intg{s}\geq 0$, $m\geq 1$, and a closed term $M$, with type
$(\liv^{\Intg{n}}_{i=1}\S\BIntT)
  \liv
  (\liv^{\Intg{s}}_{j=1}\S^{m}\BIntT)
  \liv
  \S^{m}\BIntT$, be given.
We call $\rotate{\Intg{n}}{\Intg{s}}{m}$ the closed term that takes $M$ and $\Intg{n}+\Intg{s}$ words as its arguments.
The first $\Intg{n}$ arguments can be viewed as a normal ones, while the second $\Intg{s}$ as safe.
$\rotate{\Intg{n}}{\Intg{s}}{m}$ applies $M$ to the normal arguments in the given order, while using $\BNum{s_1}$ as value at position $\Intg{s}$, shifting all the others leftward.
Namely, the \textit{behavior} is:
\[
\rotate{\Intg{n}}{\Intg{s}}{m}[M]
\BNum{n_1}\ldots\BNum{n_{\Intg{n}}}\,
\BNum{s_1}\ldots\BNum{s_{\Intg{s}}}
\rew^*
M\,\BNum{n_1}\ldots\BNum{n_{\Intg{n}}}\,
   \BNum{s_2}\ldots\BNum{s_{\Intg{s}}}\,\BNum{s_1}
\]
\par
The \textit{type} of $\rotate{\Intg{n}}{\Intg{s}}{m}$ is:
\[
((\liv^{\Intg{n}}_{i=1}\S\BIntT)
  \liv
  (\liv^{\Intg{s}}_{j=1}\S^{m}\BIntT)
  \liv
 \S^{m}\BIntT)
 \liv
(\liv^{\Intg{n}}_{i=1}\S\BIntT)
  \liv
 (\liv^{\Intg{s}}_{j=1}\S^{m}\BIntT)
 \liv
\S^{m}\BIntT
\]
\par
The \textit{definition} of $\rotate{\Intg{n}}{\Intg{s}}{m}$ is
$\bs x_1\ldots x_{\Intg{n}}\,y_{\Intg{s}}\,y_1\ldots y_{\Intg{s}-1}.
M\,x_1\ldots x_{\Intg{n}}\,y_1\ldots y_{\Intg{s}}$.
\paragraph{Multiple sharing of safe names.}
Let $\Intg{n}, \Intg{p}, \Intg{q}\geq 0$, and $ m\geq 1$.
Let $M$ be a closed term with type
$(\liv^{\Intg{n}}_{i=1}\S\BIntT)
  \liv
 (\liv^{\Intg{p}+\Intg{q}}_{j=1}\S^{m}\BIntT)
  \liv
 \S^{m}\BIntT$, when $\Intg{p}\geq 1$, and
$(\liv^{\Intg{n}}_{i=1}\S\BIntT)
 \liv
 \S^{m}\BIntT$, when $\Intg{p}= 0$.
We call $\mshare{\Intg{n}}{\Intg{p}}{\Intg{q}}{m}$ the closed term that takes $M$ and $\Intg{n}+\Intg{p}+\Intg{q}$ words as its arguments.
The first $\Intg{n}$ arguments can be viewed as normal ones, while the second $\Intg{p}+\Intg{q}$ as safe.
If $\Intg{p}>0$, and $\Intg{q}\geq 1$, then $\mshare{\Intg{n}}{\Intg{p}}{\Intg{q}}{m}$
applies $M$ to the normal and safe arguments, using $\BNum{s_{\Intg{p}}}$ as a value for the last $\Intg{q}$ safe positions.
Namely, the \textit{behavior} is:
\begin{align*}
\mshare{\Intg{n}}{\Intg{p}}{\Intg{q}}{m}[M]\,
\BNum{n_1}\ldots\BNum{n_{\Intg{n}}}\,
\BNum{s_1}\ldots\BNum{s_{\Intg{p}}}
\rew^*
&
M\,\BNum{n_1}\ldots\BNum{n_{\Intg{n}}}\,
\BNum{s_1}\ldots\BNum{s_{\Intg{p}}}
\overbrace
{\BNum{s_{\Intg{p}}}\ldots\BNum{s_{\Intg{p}}}}
^{\Intg{q}}
&
(\Intg{p}>0, \Intg{q}\geq 1)
\end{align*}
Otherwise, $\mshare{\Intg{n}}{\Intg{p}}{\Intg{q}}{m}[M]$ coincides to $M$.
\par
The \textit{type} of $\mshare{\Intg{n}}{\Intg{p}}{\Intg{q}}{m}$ is:
\begin{align*}
&
 ((\liv^{\Intg{n}}_{i=1}\S\BIntT)
   \liv
  (\liv^{\Intg{p}+\Intg{q}}_{j=1}\S^{m}\BIntT)
   \liv
  \S^{m}\BIntT)
 \liv
 (\liv^{\Intg{n}}_{i=1}\S\BIntT)
  \liv
 (\liv^{\Intg{p}}_{j=1}\S^{m+4\Intg{q}}\BIntT)
  \liv
 \S^{m+4\Intg{q}}\BIntT
 &
 (\Intg{p}\geq 1)
\\
&
 ((\liv^{\Intg{n}}_{i=1}\S\BIntT)
   \liv
  \S^{m}\BIntT)
 \liv
 (\liv^{\Intg{n}}_{i=1}\S\BIntT)
   \liv
  \S^{m}\BIntT
  &
  (\Intg{p}=0)
\end{align*}
\par
The \textit{definition} of $\mshare{\Intg{n}}{\Intg{p}}{\Intg{q}}{m}$ is:
\begin{align*}
\mshare{\Intg{n}}{\Intg{p}}{\Intg{q}}{m}[M]
&\equiv M
&(\Intg{p}=0\text{ or }\Intg{q}=0)
\\
\mshare{\Intg{n}}{\Intg{p}}{1}{m}[M]
&\equiv
\share{\Intg{n}}{\Intg{p}}{m}
      [M]
&(\Intg{p}>0)
\\
\mshare{\Intg{n}}{\Intg{p}}{\Intg{q}}{m}[M]
&\equiv
\share{\Intg{n}}{\Intg{p}}{m+4(\Intg{q}-1)}
      [\mshare{\Intg{n}}{\Intg{p}}{\Intg{q}-1}{m}[M]]
&(\Intg{p}>0, \Intg{q}>1)
\end{align*}
\paragraph{Multiple sharing and rotation of safe names.}
Let $\Intg{n}, \Intg{p}, \Intg{q}\geq 0$, and $ m\geq 1$.
Let $M$ be a closed term with type
$(\liv^{\Intg{n}}_{i=1}\S\BIntT)
  \liv
 (\liv^{\Intg{p}+\Intg{q}}_{j=1}\S^{m}\BIntT)
  \liv
 \S^{m}\BIntT$, when $\Intg{p}\geq 1$, and
$(\liv^{\Intg{n}}_{i=1}\S\BIntT)
 \liv
 \S^{m}\BIntT$, when $\Intg{p}= 0$.
We call $\rmshare{\Intg{n}}{\Intg{p}}{\Intg{q}}{m}$ the closed term that takes $M$ and $\Intg{n}+\Intg{p}+\Intg{q}$ words as its arguments.
The first $\Intg{n}$ arguments can be viewed as normal ones, while the last $\Intg{p}+\Intg{q}$ as safe ones.
If $\Intg{p}\geq 1$, then
$\rmshare{\Intg{n}}{\Intg{p}}{\Intg{q}}{m}$ applies $M$ to the normal arguments in the given order, while using $\BNum{s_1}$ as value in the last $\Intg{q}$ positions, shifting all the others leftward.
Namely, the \textit{behavior} is:
\begin{align*}
\rmshare{\Intg{n}}{\Intg{p}}{\Intg{q}}{m}[M]\,
\BNum{n_1}\ldots\BNum{n_{\Intg{n}}}\,
\BNum{s_1}\,\BNum{s_{2}}\ldots\BNum{s_{\Intg{p}}}
\rew^*
&
M\,\BNum{n_1}\ldots\BNum{n_{\Intg{n}}}\,
\BNum{s_2}\ldots\BNum{s_{\Intg{p}}}
\overbrace
{\BNum{s_{1}}\ldots\BNum{s_{1}}}
^{\Intg{q}}
&(\Intg{p}\geq 1)
\end{align*}
Otherwise, with $\Intg{p}=0$, $\rmshare{\Intg{n}}{\Intg{p}}{\Intg{q}}{m}[M]$ coincides to $M$, for any $\Intg{q}$.
\par
The \textit{type} of $\rmshare{\Intg{n}}{\Intg{p}}{\Intg{q}}{m}$ is:
\begin{align*}
&
 ((\liv^{\Intg{n}}_{i=1}\S\BIntT)
   \liv
  (\liv^{\Intg{p}+\Intg{q}}_{j=1}\S^{m}\BIntT)
   \liv
  \S^{m}\BIntT)
 \liv
 (\liv^{\Intg{n}}_{i=1}\S\BIntT)
  \liv
 (\liv^{\Intg{p}}_{j=1}\S^{m+4\Intg{q}}\BIntT)
  \liv
 \S^{m+4\Intg{q}}\BIntT
 &
 (\Intg{p}\geq 1)
\\
&
 ((\liv^{\Intg{n}}_{i=1}\S\BIntT)
   \liv
  \S^{m}\BIntT)
 \liv
 (\liv^{\Intg{n}}_{i=1}\S\BIntT)
   \liv
  \S^{m}\BIntT
  &
  (\Intg{p}=0)
\end{align*}
\par
The \textit{definition} of
$\rmshare{\Intg{n}}{\Intg{p}}{\Intg{q}}{m}$ is:
\begin{align*}
\rmshare{\Intg{n}}{\Intg{p}}{\Intg{q}}{m}[M]
&\equiv M
&(\Intg{p}=0\text{ or }\Intg{q}=0)
\\
\rmshare{\Intg{n}}{\Intg{1}}{\Intg{q}}{m}[M]
&\equiv\
\mshare{\Intg{n}}{\Intg{1}}{\Intg{q}}{m}[M]
&(\Intg{p}=1)
\\
\rmshare{\Intg{n}}{\Intg{p}}{\Intg{q}}{m}[M]
&\equiv\
\rotate{\Intg{n}}{\Intg{p}}{m+4\Intg{q}}
[\mshare{\Intg{n}}{\Intg{p}}{\Intg{q}}{m}[M]]
&(\Intg{p}>1)
\end{align*}
\paragraph{Square composition.}
The \textit{intuitive side} first.
Let $G_1,\ldots,G_{\Intg{n}'}$ be terms that we call \textit{normal} for we think of them as functions with only normal arity $\Intg{n}$. Analogously, let $H_1,\ldots,H_{\Intg{s}'}$ be terms that we call \textit{safe} since we look at them as functions with normal arity $\Intg{n}$, and safe arity $\Intg{s}_j$, for every $1\leq j\leq \Intg{s}'$. Let $\Intg{s}=\max\{\Intg{s}_1,\ldots,\Intg{s}_{\Intg{s}'},\Intg{s}'\}$; notice that $\Intg{s}$ is determined comparing the safe arguments of every safe term and their total number $\Intg{s}'$.
The behavior of
$\sqcomp
 {\Intg{n}}
 {\Intg{s}}
 {\Intg{n}'}[F,G_1\ldots G_{\Intg{n}'},H_1\ldots H_{\Intg{s}'}]$
comprises some phases, of which we have an example of result, contained in the innermost dashed box, labeled $\sqcomp{1}{3}{1}$, of Figure~\ref{figure:square-composition-block-diagram}.
Every normal argument of
$\sqcomp
{\Intg{n}}
{\Intg{s}}
{\Intg{n}'}$ is replicated as many times as $\Intg{n}'+\Intg{s}$ so that every copy can be dispatched to normal and safe terms.
\par
Then, the term $F$ is used to generate $F'$ with $\Intg{n}'$ normal and $\Intg{s}$ safe arities. $F'$ behaves like $F$ once erased its $\Intg{s}-\Intg{s}'$ arguments. For example, the bullet aside $\etp{f}$, which plays the role of $F$ in Figure~\ref{figure:square-composition-block-diagram}, represents the extension of $F'$, with respect to $F$, that erases its third safe argument.
The generation of $H'_j$ from $H_j$, with $1\leq j\leq \Intg{s}_j$, is analogous to the one of $F'$, from $F$: if necessary, every $H'_j$ erases $\Intg{s}-\Intg{s}_j$ safe arguments.
If $F'$ has to erase $\Intg{s}-\Intg{s}'$ safe arguments, we supply them as the result of $\Intg{s}-\Intg{s}'$ fake functions that erase all of their arguments and give a word as result. In Figure~\ref{figure:square-composition-block-diagram} there is a single fake function named $\BNum{0}\!\bullet\!\bullet\bullet$, yielding $\BNum{0}$.
\par
The normal and safe arguments of $\sqcomp{\Intg{n}}{\Intg{s}}{\Intg{n}'}$ are replicated by using two different processes. The one for normal arguments is the standard eager diagonal, building every copy from scratch. Instead, every replica of a safe argument is obtained by using the above combinator that rotates and shares multiple safe values.
Once all the required copies of safe arguments are at hand, they are rearranged, and appropriately distributed to $H'_1,\ldots,H'_{\Intg{s}}$.
\par
Now, the \textit{technical side}.
Let us assume to have a set of closed terms
$F,G_1,\ldots,G_{\Intg{n}'},H_1,\ldots,H_{\Intg{s}'}$ with the following types, respectively:
\begin{align*}
& (\liv^{\Intg{n}'}_{i=1}\S\BIntT)\liv
     (\liv^{\Intg{s}'}_{j=1}\S^{m}\BIntT)\liv
     \S^{m}\BIntT
\\
& (\liv^{\Intg{n}}_{i=1}\S\BIntT)\liv\S^{m}\BIntT
& (i\in\{1,\ldots,\Intg{n}'\})
\\
& (\liv^{\Intg{n}}_{i=1}\S\BIntT)\liv
      (\liv^{\Intg{s}_j}_{k=1}\S^{m}\BIntT)\liv
      \S^{m}\BIntT
& (j\in\{1,\ldots,\Intg{s}'\})
\end{align*}
Let $\Intg{s}=\max\{\Intg{s}_1,\ldots,\Intg{s}_{\Intg{s}'},\Intg{s}'\}$.
\par
The \textit{type} of $\sqcomp{\Intg{n}}{\Intg{s}}{\Intg{n}'}$ is:
\\
\begin{align*}
&
((\liv^{\Intg{n}'}_{i=1}\S\BIntT)\liv
     (\liv^{\Intg{s}'}_{j=1}\S^{m}\BIntT)\liv
     \S^{m}\BIntT)\liv
\\
&
\qquad\qquad
(\liv^{\Intg{n}'}_{k=1}
 ((\liv^{\Intg{n}}_{i=1}\S\BIntT)\liv\S^{m}\BIntT)
)\liv
\\
&
\qquad\qquad\qquad\qquad
(\liv^{\Intg{s}'}_{j=1}
 ((\liv^{\Intg{n}}_{i=1}\S\BIntT)\liv
         (\liv^{\Intg{s}_j}_{k=1}\S^{m}\BIntT)\liv
         \S^{m}\BIntT)
)\liv
\\
&
\qquad\qquad\qquad\qquad\qquad\qquad\qquad
(\liv^{\Intg{n}}_{i=1}\S\BIntT)
 \liv
 (\liv^{\Intg{s}^2}_{k=1}\S^{2m+1}\BIntT)
 \liv
 \S^{2m+1}\BIntT
\end{align*}
\par
The \textit{definition} of $\sqcomp{\Intg{n}}{\Intg{s}}{\Intg{n}'}$ is:
\begin{align*}
&
\sqcomp{\Intg{n}}
       {\Intg{s}}
       {\Intg{n}'}
       [F,G_1,\ldots,G_{\Intg{n}'},H_1,\ldots,H_{\Intg{s}'}]
\equiv
\bs n_1\ldots n_{\Intg{n}}.
 \EEmbed{2}
        {0}{\Intg{n}+\Intg{s}}
        {G}
 (\LEmbed{1}
         {1}
         {\nabla^1_{\Intg{n}'+\Intg{s}}}\,n_1)
   \ldots(\LEmbed{1}
                 {1}
                 {\nabla^1_{\Intg{n}'+\Intg{s}}}\,n_{\Intg{n}})
\end{align*}
where:
\begin{align*}
G\equiv&
   \bs \elan x_{11}\ldots x_{\Intg{n}'1}
             y_{11}\ldots y_{\Intg{s}1}\eran
   \ldots
   \bs \elan x_{1\Intg{n}}\ldots x_{\Intg{n}'\Intg{n}}
             y_{1\Intg{n}}\ldots y_{\Intg{s}\Intg{n}}\eran.
\\
&
   \bs w_{1 1}w_{1 2}\ldots w_{1\Intg{s}}.
   \bs w_{2 1}w_{2 2}\ldots w_{2\Intg{s}}.
   \ldots\ldots.
       \bs w_{\Intg{s}1}w_{\Intg{s}2}\ldots w_{\Intg{s} \Intg{s}}.
\\
&
  \, \EEmbed{m-1}{0}{\Intg{n}'+\Intg{s}}{F'}\,
   (G_1\, x_{11} \ldots x_{1\Intg{n}})
   \ldots
   (G_{\Intg{n}'}\, x_{\Intg{n}'1} \ldots x_{\Intg{n}'\Intg{n}})
\\&\,
   (\EEmbed{m-1}{0}{\Intg{n}+\Intg{s}}{H'_{1}}\,
     (\LEmbed{1}{1}{\Coerc^{m-1}}\, y_{11})
     \ldots (\LEmbed{1}{1}{\Coerc^{m-1}}\,
                y_{1\Intg{n}})\,w_{11}\,w_{21}\ldots w_{\Intg{s}1})
\\&\,
   (\EEmbed{m-1}{0}{\Intg{n}+\Intg{s}}{H'_{2}}\,
     (\LEmbed{1}{1}{\Coerc^{m-1}}\, y_{11})
     \ldots (\LEmbed{1}{1}{\Coerc^{m-1}}\,
                y_{1\Intg{n}})\,w_{12}\,w_{22}\ldots w_{\Intg{s}2})
   \ \ldots
\\&\ldots
    (\EEmbed{m-1}{0}{\Intg{n}+\Intg{s}}{H'_{\Intg{s}}}\,
     (\LEmbed{1}{1}{\Coerc^{m-1}}\, y_{\Intg{s}1})
     \ldots (\LEmbed{1}{1}{\Coerc^{m-1}}\,
               y_{\Intg{s}\Intg{n}})\,
               w_{1\Intg{s}}\,w_{2\Intg{s}}\ldots w_{\Intg{s}\Intg{s}})
\\
F'\equiv&
\bs x_1\ldots x_{\Intg{n}'}y_1\ldots y_{\Intg{s}}.
F\, x_1\ldots x_{\Intg{n}'}y_1\ldots y_{\Intg{s}'}
\\
H'_i\equiv&
\bs z_{i1}\ldots z_{i\Intg{n}}w_{i1}\ldots w_{i\Intg{s}_i}
    w_{i\Intg{s}_i+1}\ldots w_{i\Intg{s}}.
H_i\,z_{i1}\ldots z_{i\Intg{n}}\,w_{i1}\ldots w_{i\Intg{s}_i}
\qquad\ \
(i\in\{1,\ldots,\Intg{s}'\})
\\
H'_j\equiv&
\bs z_{j1}\ldots z_{j\Intg{n}}w_{j1}\ldots w_{j\Intg{s}}.
\BNum{0}
\qquad\qquad
\qquad\qquad\qquad
\qquad\qquad\qquad\
(j\in\{\Intg{s}',\ldots,\Intg{s}-\Intg{s}'\})
\end{align*}
$G$ takes both $\Intg{n}$ copies of the $\Intg{n}'+\Intg{s}$ normal arguments, generated by the $\Intg{n}$ instances of $\LEmbed{1}{1}{\nabla^{1}_{\Intg{n}'+\Intg{s}}}$, and $\Intg{s}^2$ safe arguments. Then, it dispatches them to the terms $G_1,\ldots,G_{\Intg{n}'}$ and $H'_1,\ldots,H'_{\Intg{s}}$. As we said, every $H'_i$ takes $\Intg{s}$ safe arguments and supplies only the first $\Intg{s}_i$ to $H_i$.
\par
The \textit{behavior} is:
$$
\sqcomp{\Intg{n}}{\Intg{s}}{m}[F,G_1\ldots G_{\Intg{n}'},H_1\ldots H_{\Intg{s}'}]\,
\BNum{n}_1\ldots\BNum{n}_{\Intg{n}}\,
 \overbrace{\BNum{s}_1\ldots\BNum{s}_{1}}^{\Intg{s}}
 \ldots\ldots
 \overbrace{\BNum{s}_{\Intg{s}}\ldots\BNum{s}_{\Intg{s}}}^{\Intg{s}}
\rew^+
F\,\BNum{g_1}\ldots\BNum{g_{\Intg{n}'}}\,\BNum{h_1}\ldots\BNum{h_{\Intg{s}'}}
$$
if $G_i\,\BNum{n}_1\ldots\BNum{n}_{\Intg{n}}\rew^*\BNum{g_i}$, with $1\leq i\leq \Intg{n}'$, and
$H_j\,\BNum{n}_1\ldots\BNum{n}_{\Intg{n}}\,\BNum{s}_1\ldots\BNum{s}_{\Intg{s}}\rew^*\BNum{h_j}$, with $1\leq j\leq \Intg{s}'$.
\subsection{Multiple sharing and rotation of safe names in a square composition.}
Let $\Intg{n}, \Intg{p}, i\geq 0$, $ m\geq 1$, and a closed term $M$ with type
$(\liv^{\Intg{n}}_{i=1}\S\BIntT)
  \liv
 (\liv^{\Intg{p}^2}_{j=1}\S^{m}\BIntT)
  \liv
 \S^{m}\BIntT$ be given.
We call $\mshsqcomp{\Intg{n}}{\Intg{p}}{i}{m}$ the closed term that takes $M$ and $\Intg{n}+i+\Intg{p}\,i$ words as its arguments.
The first $\Intg{n}$ arguments can be viewed as normal ones, while the last $i+\Intg{p}\,i$ as safe ones.
Then, $\mshsqcomp{\Intg{n}}{\Intg{p}}{i}{m}[M]$ replicates every of the $i$ safe arguments $\BNum{s_{\Intg{p}-i+1}}\ldots\BNum{s_{\Intg{p}}}$ as many times as $\Intg{p}$. Finally the ``blocks''
$
\BNum{s_{1}}\ldots\BNum{s_{1}}
\ldots\ldots
\BNum{s_{\Intg{p}-i}}\ldots\BNum{s_{\Intg{p}-i}}\,
\BNum{s_{\Intg{p}-i+1}}\ldots\BNum{s_{\Intg{p}-i+1}}
\ldots\ldots
\BNum{s_{\Intg{p}}}\ldots\BNum{s_{\Intg{p}}}
$,
with $\Intg{p}$ elements each,
are used as the $\Intg{p}^2$ safe arguments of $M$.
Namely, the \textit{behavior} is:
\begin{align*}
\mshsqcomp{\Intg{n}}{\Intg{p}}{i}{m}[M]\,
\BNum{n_1}\ldots \BNum{n_{\Intg{n}}}\,
\overbrace{\BNum{s_{\Intg{p}-i+1}}\ldots\BNum{s_{\Intg{p}}}}^{i}\,
\overbrace{\BNum{s_1}\ldots\BNum{s_{1}}}^{\Intg{p}}
\ldots\ldots
\overbrace{\BNum{s_{i}}\ldots\BNum{s_{i}}}^{\Intg{p}}
\rew^*
M\,
\BNum{n_1}\ldots\BNum{n_{\Intg{n}}}\,
\overbrace{\BNum{s_1}\ldots\BNum{s_{1}}}^{\Intg{p}}
\ldots\ldots
\overbrace{\BNum{s_{\Intg{p}}}\ldots\BNum{s_{\Intg{p}}}}^{\Intg{p}}
\end{align*}
The \textit{type} of $\mshsqcomp{\Intg{n}}{\Intg{p}}{i}{m}[M]$ is:
$$
((\liv^{\Intg{n}}_{i=1}\S\BIntT)
  \liv
 (\liv^{\Intg{p}^2}_{j=1}\S^{m}\BIntT)
  \liv
 \S^{m}\BIntT)
 \liv
(\liv^{\Intg{n}}_{j=1}\S\BIntT)
 \liv
 (\liv^{i+\sum^{\Intg{p}-i}_{k=1}\Intg{p}}_{j=1}\S^{m+4(\Intg{p}-1)i}\BIntT)
 \liv
 \S^{m+4(\Intg{p}-1)i}\BIntT
$$
The \textit{definition} of $\mshsqcomp{\Intg{n}}{\Intg{p}}{i}{m}[M]$ is:
\begin{align*}
\mshsqcomp{\Intg{n}}{\Intg{p}}{i}{m}[M]
&\equiv M
&(\Intg{p}\leq 1 \text{ or } i=0)
\\
\mshsqcomp{\Intg{n}}{\Intg{p}}{1}{m}[M]
&\equiv
\,
\rmshare{\Intg{n}}{1+\sum^{\Intg{p}-1}_{k=1}\Intg{p}}{\Intg{p}-1}{m}[M]
\\
\mshsqcomp{\Intg{n}}{\Intg{p}}{i}{m}[M]
&\equiv
\,
\rmshare{\Intg{n}}{i-1+\sum^{\Intg{p}-(i-1)}_{k=1}\Intg{p}}
        {\Intg{p}}
        {m+4(\Intg{p}-1)(i-1)}
[\mshsqcomp{\Intg{n}}{\Intg{p}}{i-1}{m}[M]]
&(\Intg{p}\geq i>0)
\end{align*}

%% file: completeness.tex
\section{\SRN-completeness of \WALT}
\label{section:SRN completeness of WALT}
We extend the completeness of \WALT\ from \QlSRN\ \cite{Roversi:2007-WALT-FULL} to \SRN. The key ingredients are square composition and the multiple sharing of safe names of Section~\ref{section:The safe composition in WALT}.
\paragraph{Functions of \SRN\ into \WALT.}
We start defining a map $\etp{\ }$ from the signature $\Sigma_{\SRN}$ to terms of \WALT.
Its clauses are identical to those mapping \QlSRN\ to \WALT\ \cite{Roversi:2007-WALT-FULL}, but the one mapping the composition. of course. Here they are:
\begin{enumerate}
\item
\label{embedPRNQWALL:zero}
$\etp{\zero{0}{0}} \equiv \LEmbed{1}{0}{\BNum{0}}$, while
$\etp{\zero{k}{l}}
 \equiv\bs n_{1}\ldots n_{k}\, s_{1}\ldots s_{l}.\etp{\zero{0}{0}}$,
for every $k, l$ such that $k+l\geq 1$.

\item
\label{embedPRNQWALL:sucz}
$\etp{\sucz}\equiv\BEmbed{1}{\BSuccZ}$.

\item
\label{embedPRNQWALL:suco}
$\etp{\suco}\equiv\BEmbed{1}{\BSuccO}$.

\item
\label{embedPRNQWALL:pred}
$\etp{\pred}\equiv\BEmbed{1}{\Pred}$.

\item
\label{embedPRNQWALL:proj}
$\etp{\proj{k}{l}{i}}
 \equiv \bs x_{1}\ldots x_{k+l}.x_{i}$,
   with $1\leq i\leq k+l$.

\item
\label{embedPRNQWALL:bran}
$\etp{\bran}\equiv\bs xyz.\Branch\,x\,y\,z$.

\item
\label{embedPRNQWALL:comp}
Let
$\vdash
 \ta{\etp{f}}
    {(\liv^{k'}_{i=1}\$\BIntT)\liv
     (\liv^{l'}_{i=1}\$^m\BIntT)\liv
     \$^{m}\BIntT}$, and
$\vdash
 \ta{\etp{g_i}}
    {(\liv^{k}_{i=1}\$\BIntT)\liv
     \$^{m_i}\BIntT}$, with $i\in\{1,\ldots,k'\}$, and
$\vdash
 \ta{\etp{h_j}}
    {(\liv^{k}_{i=1}\$\BIntT)\liv
     (\liv^{l_j}_{i=1}\$^{n_j}\BIntT)\liv
     \$^{n_j}\BIntT}$, with $j\in\{1,\ldots,l'\}$.
If $p=\operatorname{max}\{m,m_1,\ldots,m_{k'},n_1,\ldots,n_{l'}\}$, and
$l=\max\{l_1,\ldots,l_{l'},l'\}$, then:
\small
\begin{align*}
&
 \etp{
 \comp{k}
      {l}
      {k'}
      {l'}
      {f,g_1,\ldots,g_{k'},h_1,\ldots,h_{l'}}
     }\equiv
\\
&\qquad
\mshsqcomp{k}{l}{l}{2p+1}
[
\sqcomp{k}{l}{k'}
   [\bs x_1\ldots x_{k'}.
    \EEmbed{p-m}{k'}{l'}{\etp{f}}
    (\LEmbed{1}{1}{\Coerc^{p-m-1}}\,x_1)\ldots(\LEmbed{1}{1}{\Coerc^{p-m-1}}\,x_{k'})
\\
&\qquad
 \phantom{\mshsqcomp{k}{l}{l}{2p+1}
          \sqcomp{k}{l}{k'}[}
   ,\bs x_1\ldots x_{k}.
    \EEmbed{p-m_1}{k}{0}{\etp{g_1}}
    (\LEmbed{1}{1}{\Coerc^{p-m_1-1}}\,x_1)\ldots(\LEmbed{1}{1}{\Coerc^{p-m_1-1}}\,x_{k})
\\
&\qquad
 \phantom{\mshsqcomp{k}{l}{l}{2p+1}
          \sqcomp{k}{l}{k'}[}
   \ldots
\\
&\qquad
 \phantom{\mshsqcomp{k}{l}{l}{2p+1}
          \sqcomp{k}{l}{k'}[}
   ,\bs x_1\ldots x_{k}.
    \EEmbed{p-m_{k'}}{k}{0}{\etp{g_{k'}}}
    (\LEmbed{1}{1}{\Coerc^{p-m_{k'}-1}}\,x_1)
    \ldots(\LEmbed{1}{1}{\Coerc^{p-m_{k'}-1}}\,x_{k})
\\
&\qquad
 \phantom{\mshsqcomp{k}{l}{l}{2p+1}
          \sqcomp{k}{l}{k'}[}
   ,\bs x_1\ldots x_{k}.
    \EEmbed{p-n_1}{k}{l_1}{\etp{h_1}}
    (\LEmbed{1}{1}{\Coerc^{p-n_1-1}}\,x_1)\ldots(\LEmbed{1}{1}{\Coerc^{p-n_1-1}}\,x_{k})
\\
&\qquad
 \phantom{\mshsqcomp{k}{l}{l}{2p+1}
          \sqcomp{k}{l}{k'}[}
   \ldots
\\
&\qquad
 \phantom{\mshsqcomp{k}{l}{l}{2p+1}
          \sqcomp{k}{l}{k'}[}
   ,\bs x_1\ldots x_{k}.
    \EEmbed{p-n_{l'}}{k}{l_{l'}}{\etp{h_{l'}}}
    (\LEmbed{1}{1}{\Coerc^{p-n_{l'}-1}}\,x_1)
    \ldots(\LEmbed{1}{1}{\Coerc^{p-n_{l'}-1}}\,x_{k})
]
]
\enspace .
\end{align*}
\normalsize

\item
\label{embedPRNQWALL:rec}
If
$\vdash
 \ta{\etp{f_i}}
    {\$\BIntT\liv
     (\liv^{k}_{i=1}\$\BIntT)\liv
     (\liv^{l}_{i=1}\$^{m_i}\BIntT)\liv
     \$^{m_i}\BIntT\liv
     \$^{m_i}\BIntT}$, with $i\in\{0,1\}$, and
$\vdash
 \ta{\etp{g}}
    {(\liv^{k}_{i=1}\$\BIntT)\liv
     (\liv^{l}_{i=1}\$^{m}\BIntT)\liv
     \$^{m}\BIntT}$,
then:
\small
$$
\etp{\rec{k+1}{l}{g,f_0,f_1}}\equiv
\Iter{1+\Intg{k}}{l}{F_0}{F_1}{G}
\enspace ,
$$
\normalsize
where
$G\equiv\EEmbed{p-m}
          {k+1}
          {l+1}
          {\bs n_0\,n_1\ldots n_{k}\,s_1\ldots s_{l}\,r.
           \etp{g}\, n_1\ldots n_k\,s_1\ldots s_l}
$,
$F_i\equiv\EEmbed{p-m_i}
            {k+1}
            {l+1}
            {\etp{f_i}}$,
with $p=\operatorname{max}\{m_0,m_1,m\}$, and $i\in\{0,1\}$.
\end{enumerate}
\textbf{Interpreting \SRN\ to \WALT.}
\label{definition:Interpreting SRN to WALT}
Let $\mathcal R$ be the set of environments, such that, every $\rho\in{\mathcal R}$ is a map from $\SRNVnames$ to $\Nat$. Then, $\srtw{\ }{}$ is a map from a pair in
$(\SRN\cup\Sigma_{\SRN})\times{\mathcal R}$,
to \WALT, inductively defined on its first argument:
\small
\begin{align*}
\srtw{x}{\rho}&=\srtw{\rho(x)}{\rho}\qquad\qquad\qquad\qquad\qquad\qquad\qquad
(x\in\SRNVnames)\\
\srtw{0}{\rho}&=\etp{0}\\
\srtw{f}{\rho}&=\etp{f}\quad\qquad\qquad\qquad\qquad\qquad\qquad\qquad
(f\in\Sigma_{\SRN})\\
\srtw{f(t_1,\ldots, t_k, u_1,\ldots, u_l)}{\rho}
&=
\\
&
\vspace{-1cm}
\EEmbed{v-u+1-m}
       {0}
       {l}
       {\EEmbed{u-1}{0}{k+l}{\srtw{f}{\rho}}
        (\LEmbed{u-p_{1}}{0}{\srtw{t_1}{\rho}})
        \ldots
        (\LEmbed{u-p_{k}}{0}{\srtw{t_k}{\rho}})
       }
\\
&
\qquad\ \ \
(\LEmbed{v-q_{1}}{0}{\srtw{u_1}{\rho}})
\ldots
(\LEmbed{v-q_{l}}{0}{\srtw{u_l}{\rho}})
\qquad
(f\in\Sigma^{k,l}_{\SRN})
\end{align*}
\normalsize
when $u=\max\{m,p_1,\ldots,p_k\}$, $v=\max\{u-1+m,q_1\ldots,q_l\}$, and:
\small
\begin{align*}
&\vdash
 \ta{\srtw{f}{\rho}}
    {(\liv^{k}_{i=1}\$\BIntT)\liv
     (\liv^{l}_{j=1}\$^m\BIntT)\liv
     \$^m\BIntT}\\
&\vdash
 \ta{\srtw{t_i}{\rho}}
    {\$^{p_i}\BIntT}& i\in\{1,\ldots,k\}
\phantom{\enspace .}
\\
&\vdash
 \ta{\srtw{u_j}{\rho}}
    {\$^{q_j}\BIntT}& j\in\{1,\ldots,l\}
\enspace .
\end{align*}
\normalsize
Otherwise, $\srtw{\ }{}$ is undefined.
\par
\textbf{Weight of a term in \SRN.}
\label{definition:Weight of a term in SRN}
For proving the statement that formalizes how we can embed \SRN\ into \WALT\ (Theorem~\ref{theorem:SRN is a subsystem of WALT} below)
we need a notion of weight of a \textit{closed} term in \SRN, which, essentially, gives a measure of its impredicativity.
For every \textit{closed} term $t\in\SRN\cup\Sigma_{\SRN}$, $\wght{t}{}$ is the \textit{weight of $t$}, defined by induction on $t$. If $t$ is one among zero, predecessor, successor, projection, and branching, then $\wght{t}{}=0$.
Otherwise:
\small
\begin{align*}
\wght{
\comp{k}
     {l}
     {k'}
     {l'}
     {f,g_1,\ldots,g_{k'},h_1,\ldots,h_{l'}}
}{}
&=
3\max\{
\wght{f}{},
\wght{g_1}{}
,\ldots,
\wght{g_k}{},
\wght{h_1}{}
,\ldots,
\wght{h_l}{}
,\frac{1}{3}\}
\\
\wght{
\rec{k+1}{l}{g,h_0,h_1}
}{}
&=
2\max\{
\wght{g}{},
\wght{h_0}{},
\wght{h_1}{}
,\frac{1}{2}
\}
\\
\wght{f(t_1, \ldots, t_k, u_1, \ldots, u_l)}{}
&=
2\max\{
\wght{f}{},
\wght{t_1}{}
,\ldots,
\wght{t_k}{},
\wght{u_1}{}
,\ldots,
\wght{u_l}{}
,\frac{1}{2}\}
\end{align*}
\normalsize
\vspace{-.5cm}
\begin{theorem}[\SRN\ is a subsystem of \WALT.]
\label{theorem:SRN is a subsystem of WALT}
Let $k, l\in\Nat$, $f\in\Sigma^{k,l}_{\SRN}$, and $t, t_1, \ldots, t_k, u_1, \ldots,u_l$ be terms of \SRN.
\begin{enumerate}
\item
\label{theo:realizPRNQ1}
There is an $m\geq 1$ such that
$\vdash
 \ta{\etp{f}}
    {(\liv_{i=1}^{k}\$\BIntT)\liv
     (\liv_{j=1}^{l}\$^{m}\BIntT)\liv
     \$^{m}\BIntT}
$.

\item
\label{theo:realizPRNQ2}
$\srtw{f(t_1, \ldots, t_k, u_1, \ldots, u_l)}{\rho}$ is defined,
for every $\rho$.

\item
\label{theo:realizPRNQ3}
$\vdash
 \ta{\srtw{t}{}}
    {\$^{m}\BIntT}$ with $m\leq \wght{t}{}$.

\item
\label{theo:realizPRNQ4}
$\srtw{n}{}\rew^+ \BNum{n}$, for every $n\geq 0$.

\item
\label{theo:realizPRNQ5}
If $f(n_1,\ldots,n_k,s_1,\ldots,s_l)=n$, then
$\srtw{f(n_1,\ldots,n_k,s_1,\ldots,s_l)}{}\rew^*\BNum{n}$,
for every $n_1,\ldots,n_k,s_1,\ldots,$ $s_l,n\in\Nat$.
\end{enumerate}
\end{theorem}
Point~\ref{theo:realizPRNQ1} is a direct consequence of the typing of the combinators of \WALT\ that we use in the definition of $\etp{f}$.
Point~\ref{theo:realizPRNQ2} follows from point~\ref{theo:realizPRNQ1} here above and from the definition of $\srtw{\ }{}$.
Point~\ref{theo:realizPRNQ3} holds by induction on $t$.
Point~\ref{theo:realizPRNQ4} holds by induction on $n$.
Point~\ref{theo:realizPRNQ5} holds by induction on $f$.
Finally, by structural induction on $t$, we have:
\begin{corollary}[The embedding of \SRN\ into \WALT\ is sound.]
\label{corollary:The embedding of SRN into WALT is sound}
Let $t\in\SRN$, and $n\in\Nat$.
If $t=n$, then $\srtw{t}{\rho}\rew^{+}\BNum{n}$, for every environment $\rho$.
\end{corollary}

%% file: conclusions.tex
\section{Conclusions and future work}
\label{section:Conclusions and future work}
\WALT\ is the first higher-order deductive system, derived from Linear logic, such that: (i) is sound and complete w.r.t. \FP, (ii) is complete w.r.t. \SRN, (iii) makes evident the layered nature of the almost flat normal/safe hierarchy about the arguments of the terms of \SRN, and (iv) no constant symbol is required to obtain the point (iii), since every datatype can defined from scratch.
\par
In particular, point (ii) allows to say that the less an argument of a term of \SRN\ is ``polynomially impredicative'', the deeper its representation is inside the stratified structure of the derivations of \WALT. This relation between the polynomial impredicativity and the stratification suggests that a relation between \WALT\ and Higher type ramified recurrence (\BNS) \cite{Bellantoni00APAL,Bellantoni01MSS}, or Higher linear ramified recursion (\HOLRR) \cite{DalLago+Martini+Roversi:2004-TYPES} should exist. We think that the most intriguing is the one between \BNS\ and \WALT. The reason is that \BNS\ characterizes \FP\ by a careful interplay of conditions about its types, built on an almost linear arrow type and $!$-modal types, and its terms, derived from G\"{o}del \ST\ \cite{Godel:1958-T}. The notions of complete/incomplete types, linked to their modality, the possibility of duplicating at will only ground types, and the affinability, which expresses linearity constraints on the bound variables of incomplete types, strongly recall the properties we enforce on $\liv$, on its arguments and on the $\$$-modal assumptions of $!$-boxes in \WALT.
\par
A further investigation could go in the ``backward'' direction, namely from the structural proof-theoretical world, represented by \WALT, to the recursive theoretical one, represented by \SRN.
\par
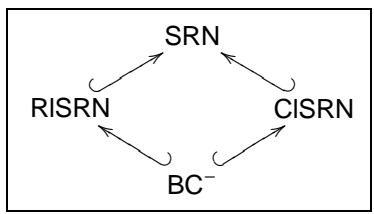
\begin{figure}[ht]
\begin{center}
\fbox{
\xymatrix@=.5cm{
& \SRN &
\\
\RlSRN \ar@{^{(}->}[ur] & & \QlSRN \ar@{_{(}->}[ul]
\\
& \BC^- \ar@{_{(}->}[ul] \ar@{^{(}->}[ur] &
}
}
\end{center}
\caption{A simple syntactic hierarchy}
\label{figure:simple SRN hierarchy}
\end{figure}

Let us look at Figure~\ref{figure:simple SRN hierarchy}.
It fixes a hierarchy, based on syntactic restrictions.
We already know what \SRN, \QlSRN, and \BC$^-$ are.
Instead, \RlSRN, called \textit{Recursion-linear} \SRN, is ``new''. It is defined as the ``complement'' of \QlSRN\ w.r.t. \SRN\ by \textit{restricting the recursive scheme} of \SRN\ to one that uses its safe variables linearly, while leaving the composition scheme untouched.
\par
\SRN\ and \QlSRN\ should be both polytime complete, as consequence of the moral equivalence ``full composition scheme of \SRN\ $\simeq$ recursion scheme + linear composition scheme of \QlSRN'', we have proved in this work.
\par
Moreover, we know that \BC$^-$ is contained into the class of deterministic logarithmic space \cite{Neergaard:APLAS-04}. We can ask which is the space complexity of \QlSRN, which should not coincide to the one of \SRN, because they develop different computation processes of, very likely, \FP. Of course, the same questions may be asked and answered about \RlSRN, so inducing a space hierarchy, which originates from a syntactic analysis of \SRN, in its turn coming from the structural proof-theoretic roots of \WALT.

%% file: proofs.tex
\section{Some detailed proofs}
\subsection{$\mshsqcomp{\Intg{n}}{\Intg{p}}{i}{m}[M]$ is well typed.}
For every $\Intg{p}$, this can be proved by cases on the value of $\Intg{p}$, and by induction on $i$.
When $\Intg{p}\leq 1$ or $i=0$, $\mshsqcomp{\Intg{n}}{\Intg{p}}{i}{m}[M]$ is $M$ with, at most, a single safe argument. So, the statement trivially holds.
Let us assume $\Intg{p}>1$ and $\Intg{p}\geq i\geq 1$.
\par
The base case has $i=1$.
We start from the type
$(\liv^{\Intg{n}}_{i=1}\S\BIntT)
  \liv
 (\liv^{\Intg{p}^2}_{j=1}\S^{m}\BIntT)
  \liv
 \S^{m}\BIntT$ of $M$, observing that:
\[
\Intg{p}^2
=\sum^{\Intg{p}}_{k=1}\Intg{p}
=\Intg{p}+\sum^{\Intg{p}-1}_{k=1}\Intg{p}
=\Intg{p}-1+1+\sum^{\Intg{p}-1}_{k=1}\Intg{p}
=(1+\sum^{\Intg{p}-1}_{k=1}\Intg{p})+(\Intg{p}-1)
\]
So, we can apply the clause defining
$\mshsqcomp{\Intg{n}}{\Intg{p}}{1}{m}[M]$, getting that its type is the one of
$$
\rmshare{\Intg{n}}{1+\sum^{\Intg{p}-1}_{k=1}\Intg{p}}{\Intg{p}-1}{m}[M]
\enspace,
$$
namely
$(\liv^{\Intg{n}}_{j=1}\S\BIntT)
 \liv
 (\liv^{1+\sum^{\Intg{p}-1}_{k=1}\Intg{p}}_{j=1}\S^{m+4(\Intg{p}-1)}\BIntT)
 \liv
 \S^{m+4(\Intg{p}-1)}\BIntT
$.
\par
By induction, the type of
$\mshsqcomp{\Intg{n}}{\Intg{p}}{i-1}{m}[M]$ is
$
(\liv^{\Intg{n}}_{j=1}\S\BIntT)
 \liv
 (\liv^{(i-1)+\sum^{\Intg{p}-(i-1)}_{k=1}\Intg{p}}_{j=1}\S^{m+4(\Intg{p}-1)(i-1)}\BIntT)
 \liv
 \S^{m+4(\Intg{p}-1)(i-1)}\BIntT
$. Observing that the following of equivalences hold:
\[
(i-1)+\sum^{\Intg{p}-(i-1)}_{k=1}\Intg{p} =
(i-1)+\sum^{\Intg{p}-i+1}_{k=1}\Intg{p} =
(i-1)+\Intg{p}+\sum^{\Intg{p}-i}_{k=1}\Intg{p} =
i-1+\Intg{p}-1+1+\sum^{\Intg{p}-i}_{k=1}\Intg{p} =
(i+\sum^{\Intg{p}-i}_{k=1}\Intg{p})+(\Intg{p}-1)
\]
we can transform the type of
$\mshsqcomp{\Intg{n}}{\Intg{p}}{i-1}{m}[M]$ so that we can use it as argument of $\rmshare{\Intg{n}}{i+\sum^{\Intg{p}-i}_{k=1}}{p-1}{m+4(\Intg{p}-1)(i-1)}$. By definition, we get a term with the type we need.
\subsection{$\mshsqcomp{\Intg{n}}{\Intg{p}}{i}{m}[M]$ well behaves.}
Let $\Intg{p}=0$. Then,
\begin{align}
\mshsqcomp{\Intg{n}}{\Intg{0}}{i}{m}[M]\,
\BNum{n_1}\ldots \BNum{n_{\Intg{n}}}\,
\overbrace{\BNum{s_{\Intg{0}-i+1}}\ldots\BNum{s_{\Intg{0}}}}^{i}\,
\overbrace{\BNum{s_1}\ldots\BNum{s_{1}}}^{\Intg{0}}
\ldots\ldots
\overbrace{\BNum{s_{i}}\ldots\BNum{s_{i}}}^{\Intg{0}}
\equiv
M\,
\BNum{n_1}\ldots\BNum{n_{\Intg{n}}}
\label{align-mshsqcomp-p=0}
\end{align}
where the sequences of safe arguments cannot exist since we assume that the indices of the safe arguments start from $1$. So, \eqref{align-mshsqcomp-p=0} rewrites to $M\,\BNum{n_1}\ldots\BNum{n_{\Intg{n}}}$ in $0$ steps.
\par
Let $\Intg{p}\geq 1$ and $i=0$. Then,
\begin{align}
\mshsqcomp{\Intg{n}}{\Intg{p}}{0}{m}[M]\,
\BNum{n_1}\ldots \BNum{n_{\Intg{n}}}\,
\overbrace{\BNum{s_{\Intg{p}-0+1}}\ldots\BNum{s_{\Intg{p}}}}^{0}\,
\overbrace{\BNum{s_1}\ldots\BNum{s_{1}}}^{\Intg{p}}
\ldots\ldots
\overbrace{\BNum{s_{\Intg{p}-0}}\ldots\BNum{s_{\Intg{p}-0}}}^{\Intg{p}}
\equiv
M\,
\BNum{n_1}\ldots\BNum{n_{\Intg{n}}}
\overbrace{\BNum{s_1}\ldots\BNum{s_{1}}}^{\Intg{p}}
\ldots\ldots
\overbrace{\BNum{s_{\Intg{p}}}\ldots\BNum{s_{\Intg{p}}}}^{\Intg{p}}
\label{align-mshsqcomp-i=0}
\end{align}
where the inital sequences of safe arguments are those required directly by $M$. This is why $\mshsqcomp{\Intg{n}}{\Intg{p}}{0}{m}[M]$ coincides to $M$ and the statement holds relatively \eqref{align-mshsqcomp-i=0}.
\par
Let $\Intg{p}\geq 1$ and $i>0$.
By induction, we have:
\begin{align*}
\mshsqcomp{\Intg{n}}{\Intg{p}}{i-1}{m}[M]\,
\BNum{n_1}\ldots \BNum{n_{\Intg{n}}}\,
\overbrace{\BNum{s_{\Intg{p}-(i-1)+1}}\ldots\BNum{s_{\Intg{p}}}}^{i-1}\,
\overbrace{\BNum{s_1}\ldots\BNum{s_{1}}}^{\Intg{p}}
\ldots\ldots
\overbrace{\BNum{s_{i-1}}\ldots\BNum{s_{i-1}}}^{\Intg{p}}
\rew^*
M\,
\BNum{n_1}\ldots\BNum{n_{\Intg{n}}}\,
\overbrace{\BNum{s_1}\ldots\BNum{s_{1}}}^{\Intg{p}}
\ldots\ldots
\overbrace{\BNum{s_{\Intg{p}}}\ldots\BNum{s_{\Intg{p}}}}^{\Intg{p}}
\end{align*}
We also know that, for every term $N$ with the right type, depending on $m'$:
\begin{align}
\nonumber
&
\rmshare{\Intg{n}}{i-1+\sum_{k=1}^{\Intg{p}-(i-1)}\Intg{p}}{\Intg{p}}{m'}[N]\,
\BNum{n_1}\ldots\BNum{n_{\Intg{n}}}\,
\overbrace{\BNum{s_{\Intg{p}-(i-1)}}\,
           \BNum{s_{\Intg{p}-(i-1)+1}}\ldots\BNum{s_{\Intg{p}}}}^{i-1}\,
\overbrace{\BNum{s_1}\ldots\BNum{s_{1}}}^{\Intg{p}}
\ldots\ldots
\overbrace{\BNum{s_{\Intg{p}-(i-1)-1}}\ldots\BNum{s_{\Intg{p}-(i-1)-1}}}^{\Intg{p}}
\\
\label{align-rmshare-generic-N}
&\rew^*
N\,\BNum{n_1}\ldots\BNum{n_{\Intg{n}}}\,
\overbrace{\BNum{s_{\Intg{p}-(i-1)+1}}\ldots\BNum{s_{\Intg{p}}}}^{i-1}\,
\overbrace{\BNum{s_1}\ldots\BNum{s_{1}}}^{\Intg{p}}
\ldots\ldots
\overbrace{\BNum{s_{\Intg{p}-(i-1)-1}}\ldots\BNum{s_{\Intg{p}-(i-1)-1}}}^{\Intg{p}}
\overbrace{\BNum{s_{\Intg{p}-(i-1)}}\ldots\BNum{s_{\Intg{p}-(i-1)}}}^{\Intg{p}}
\end{align}
So, in \eqref{align-rmshare-generic-N}, $N$ can be $\mshsqcomp{\Intg{n}}{\Intg{p}}{i-1}{m}[M]$ with
$m'=m+4(\Intg{p}-1)(i-1)$. But, by definition,
$$\rmshare{\Intg{n}}
         {i-1+\sum_{k=1}^{\Intg{p}-(i-1)}\Intg{p}}
         {\Intg{p}}
         {m+4(\Intg{p}-1)(i-1)}
[\mshsqcomp{\Intg{n}}{\Intg{p}}{i-1}{m}[M]]
$$
is
$\rmshare{\Intg{n}}{i+\sum_{k=1}^{\Intg{p}-i}\Intg{p}}{\Intg{p}}{m}[M]$, hence with the behaviour we want.